\documentclass[%
 reprint,
nofootinbib,
 amsmath,amssymb,
 aps,
]{revtex4-2}
\usepackage{hyperref}
\renewcommand{\Re}{\operatorname{Re}}
\renewcommand{\Im}{\operatorname{Im}}
\usepackage{isotope}
\usepackage{bbm}
\usepackage{xcolor}
\def\Ars{$\isotope[40]{A\lowercase{r}}$}

\def\Ar40{\Ars}
\usepackage{epsfig}
\usepackage{graphicx}
\usepackage{dcolumn}
\usepackage{bm}
\usepackage{isotope}

\begin{document}
\newcommand{\goo}{\,\raisebox{-.5ex}{$\stackrel{>}{\scriptstyle\sim}$}\,}
\newcommand{\loo}{\,\raisebox{-.5ex}{$\stackrel{<}{\scriptstyle\sim}$}\,}


\title{A New Model of Intranuclear Neutron-Antineutron Transformations in \isotope[16][8]{O}}

\author{J. L. Barrow}
 \email{jbarrow@fnal.gov}
 \affiliation{The Massachusetts Institute of Technology, Department of Physics, 77 Massachusetts Avenue, Building 4, Room 304, Cambridge, MA 02139, USA}
 \altaffiliation[Also at ]{Tel Aviv University}
 \altaffiliation{formerly at The University of Tennessee, Knoxville.}

\author{A. S. Botvina}
 \email{botvina@inr.ru}
 \affiliation{Institute for Nuclear Research, Russian Academy of Sciences, Prospekt 60-letiya Oktyabrya 7a, 117312 Moscow, Russia}

\author{E. S. Golubeva}
 \email{golubeva@inr.ru}
 \affiliation{Institute for Nuclear Research, Russian Academy of Sciences, Prospekt 60-letiya Oktyabrya 7a, 117312 Moscow, Russia}

\author{J.-M. Richard}
 \email{j-m.richard@ipnl.in2p3.fr}
 \affiliation{Institut de Physique des 2 Infinis de Lyon, Universit\'e de Lyon, CNRS-IN2P3--UCBL, 4 rue Enrico Fermi, Villeurbanne, France}

\date{\today}

\begin{abstract}
There has been much work in recent years pertaining to viability studies for the intranuclear observation of neutron-antineutron transformations. These studies begin firstly with the design and implementation of an event generator for the simulation of this rare process, where one hopes to retain as much of the underlying nuclear physics as possible in the initial state, and then studying how these effects may perturb the final state observable particles for detector efficiency studies following simulated reconstruction. There have been several searches for intranuclear neutron-antineutron transformations, primarily utilizing the \isotope[16][8]{O} nucleus, and completed within large underground water Cherenkov detectors such as Super-Kamiokande. The latest iteration of a generator is presented here for use in such an experiment. This generator includes several new features, including a new radial (position) annihilation probability distribution and related intranuclear suppression factor for \isotope[16][8]{O}, as well as a highly general, modern nuclear multifragmentation model with photonic de-excitations. The latter of these may allow for improved identification of the signal using large underground detectors such as Super-Kamiokande and the future Hyper-Kamiokande, potentially increasing the overall signal efficiencies of these rare searches. However, it should be noted that certain fast photonic de-excitations may be washed out by $\pi^0$ decays to photons. These new features implemented in these $\bar{n}$\isotope[15][8]{O} simulations increase the overall physical realism of the model, and are easily portable to other future searches such as to extranuclear $\bar{n}$\isotope[12][6]{C} for the ESS NNBAR experiment, as well as intranuclear $\bar{n}$\isotope[39][18]{Ar} used in DUNE.
\end{abstract}

\maketitle

\section{Introduction} \label{sec:intro}
Baryon number ($\mathcal{B})$ must be violated via the Sakharov conditions~\citep{Sakharov:1967dj} in order to explain the baryon asymmetry of the universe. This is the case despite the fact that within the Standard Model (SM) $\mathcal{B}$ is accidentally conserved~\citep{GonzalezGarcia:2002dz} at perturbative scales, and is only infinitesimally violated within nonpertubative regimes~\citep{tHooft:1976snw}. Even though it is a key beyond Standard Model (BSM) prediction, it has yet to be observed through the now classic $|\Delta\mathcal{B}|=1$ proton decay~\citep{Takenaka:2020vqy}. Furthermore, it has been shown that many BSM theories do not require such decays~\citep{Arnold:2012sd,Nussinov:2001rb}, and many actually prefer $|\Delta\mathcal{B}|=2$ operators~\citep{Babu:2020nnh} such as those associated with dinucleon decays and neutron-antineutron transformations ($n\rightarrow\bar{n}$)~\citep{Kuzmin:1970nx,Kuzmin:1987wn,Babu:2006xc,Babu:2013yca}. It may also be said that, generically, only $\mathcal{B-L}$-violating processes may contribute to the baryon abundance~\citep{Dolgov:1991fr,Babu:2020nnh}, as all primordial $\mathcal{B}$ asymmetries originating from higher scale interactions are usually washed out via sphalerons; this is not the case in models of leptogenesis~\citep{Fukugita:1986hr,Pascoli:2016gkf,Chun:2017spz}.

Post-sphaleron baryogenesis~\citep{Babu:2006xc,Babu:2013yca} provides a testable framework through the potential observation of $n\rightarrow\bar{n}$, predicting both lower and upper limits to the characteristic mean \textit{free} transformation (oscillation) time, $\tau_{n\bar{n}}$. Any observation of $n\rightarrow\bar{n}$ by one instrument, be it in an intranuclear or extranuclear context, requires confirmation studies in complimentary experiments using different techniques (such as with a free $n$ beam) or within different nuclei. The relation of the mean intranuclear transformation time, $\tau_M$, to $\tau_{n\bar{n}}$ is  $\tau_M=T_{R}\,\tau_{n\bar{n}}^2$ and requires the computation of an intranuclear suppression factor, $T_{R}$; deviations from these relations may also yield signs of new physics~\citep{Berezhiani:2015afa}, though of course require an initial observation above backgrounds. 

A series of recent workshops and associated reports have served as both a nursery and a stage for much of this collective progress, and have inspired this work; for those interested, consider~\citep{Babu:2020nnh,Young:2019pzq,Addazi:2020nlz,Barrow:2019viz,Oosterhof:2019dlo,Rinaldi:2018osy,Rinaldi:2019thf,Abe:2020ywm,Girmohanta:2020eav,Girmohanta:2020qfd,Abi:2020kei,Abi:2020evt}.

The work discussed here is aimed at creating a model for describing the $n\rightarrow\bar{n}$ transition's initial position within the nucleus, along with its subsequent annihilation on a constituent nucleon within \isotope[16][8]{O}. Such a generator is useful to large underground water Cherenkov detectors such as the Super-Kamiokande experiment. Within detectors such as these, reconstruction of nuclear remnant de-excitation photons are in principle possible at appropriate energies, and current searches for $\mathcal{B}$-violation in Super-Kamiokande~\citep{Takenaka:2020vqy} include these de-excitations within their simulations. Given these distinctive experimental capabilities, the inclusion of simultaneous break-up and associated $\gamma$ de-excitations have been added to this version of the model, improving on the de-excitation simulations already implemented in previous iterations; this also increases the overall physical realism of the generator.

This paper is organized as follows: Sec.~\ref{sec:review} will shortly summarize the main features of the latest model, and will highlight some of the main contrasts between this work and the model currently used within Super-Kamiokande~\citep{Abe:2020ywm}. In Sec.~\ref{sec:nuceffects}, a new calculation describing the transition and subsequent annihilation of an $\bar{n}$ in \isotope[16][8]{O} is considered, permitting the computation of its associated intranuclear suppression factor. In Sec.~\ref{sec:SMM}, a newly added statistical model for the de-excitation of nuclear remnants resulting from the annihilation is described, taking into account the emission of photons. Sec.~\ref{sec:validation} will present some details for the simulation of intranuclear $n\rightarrow\bar{n}$ in \isotope[16][8]{O} most useful for large underground Cherenkov detectors such as Super-Kamiokande and the future Hyper-Kamiokande, 
followed by some conclusions.

\section{Review of the Monte Carlo model} \label{sec:review}

\subsection{The current generator} \label{sec:current_generator}
Over the last several years, the core members of this working group have improved the underlying physical modeling of both extranuclear and intranuclear $\bar{n}$ annihilation following an $n\rightarrow\bar{n}$ transition~\citep{Golubeva:2018mrz,Barrow:2019viz}. Consider now some of the main features of this model and some important differences with the Monte Carlo generator currently used by Super-Kamiokande (MCSK).
\begin{enumerate}
    \item The initial annihilation position is taken from theoretical calculations for both extranuclear and intranuclear annihilations; examples of this past work include extranuclear $\bar{p}$\isotope[12][6]{C} and $\bar{n}$\isotope[12][6]{C}~\citep{Golubeva:2018mrz} (the former of which has been compared to available data), and for the intranuclear case in $\bar{n}$\isotope[39][18]{Ar}~\citep{Barrow:2019viz}. The calculation of the radial dependence of the annihilation position for \isotope[16][8]{O} will be presented in Sec.~\ref{sec:nuceffects}. In all cases, calculations indicate the predominately peripheral nature of the annihilation, which affects the number of final state interactions (FSIs) for the simulation. Within the current MCSK, and in contrast to the model discussed here, it is assumed that the annihilation occurs with equal probability over the entire volume of the nucleus~\citep{Abe:2020ywm}.
    
    \item  The Monte Carlo generator describes the nucleus as a local, degenerate Fermi gas of nucleons enclosed within a spherical potential well with a radius equal to the nuclear radius. To take into account the diffuse boundary of the nucleus, the nuclear density distribution is split into seven concentric zones\footnote{In addition, an eighth zone at high radii beyond the periphery takes on an an extremely low density.}, within which the nucleon density is considered to be constant. The momentum distribution of the nucleons in individual zones will be the same as for a degenerate Fermi gas, although corresponding to an $i$th-zone’s boundary Fermi momentum value~\citep{Golubeva:2018mrz}. Thus, in this model, there is a correlation between the radius and momentum of intranuclear nucleons, as the Fermi boundary energy is higher the larger the density of the nuclear medium, and so is local in character. In the MCSK, the Fermi motion is simulated via a nonlocal spectral function sourced directly from experimental measurements~\citep{Abe:2020ywm}.
    
    \item The phenomena of $\bar{N}N$ annihilation can lead to the creation of many particles through many possible (at times $\sim200$) exclusive reaction channels~\citep{Golubeva:2018mrz}. Many neutral particles may be present, which can make experimental study quite difficult, and so experimental information for exclusive channels is known only for a small fraction of possible annihilation channels. For this reason, semi-empirical tables of annihilation channels are employed for use in the modeling of the annihilation. These are obtained as follows: First, all experimentally measured channels are included. Then, by using isotopic relations, probabilities were found for those channels which have the same configurations but different particle charges. Finally, the predictions of a statistical model with SU(3) symmetry produces the remaining intermediate channels. It is considered that channels for $\bar{n}n$ are identical to $\bar{p}p$ channels, and that annihilation channels for $\bar{n}p$ are charge conjugated to $\bar{p}n$ channels~\citep{Golubeva:2018mrz,Barrow:2019viz}. The $\bar{p}p$ simulation results of this model are compared with experimental data on $\bar{p}p$ annihilation at rest, where Table~\ref{tab:newmesonmultiplicities} shows the average multiplicity of mesons formed therein. The simulation results are within the range of experimental uncertainties. From these simulation results, it follows that more than $35\%$ of all pions have been formed by the decay of heavy mesonic resonances. The decays of $\omega$ and $\eta$ mesons also act as sources of high-energy photons. In the current MCSK, a smaller (but greatly expanded~\citep{Abe:2011ky}) portion of these possible annihilation channels~\citep{Abe:2020ywm} are utilized.
    
    \item The propagation and interaction of annihilation mesons in the nuclear medium are simulated in great detail within the Intranuclear Cascade (INC) model~\citep{Golubeva:2018mrz}. The model similarly includes effects related to the influence of the nuclear environment via the introduction of an antinucleon potential and ``off-shell" masses for both the $\bar{n}$ and annihilation partner nucleon~\citep{Barrow:2019viz}. This approach demonstrates a good description of the few available experimental data on $\bar{p}A$ annihilation at rest~\citep{Golubeva:2018mrz}. In Table~\ref{tab:multiplicities-exper_calc}, the experimental multiplicities of the emitted final state annihilation-generated pions and the energy carried away by those pions and photons (from heavy resonance decays) are shown. It can be seen that the simulation results are in good agreement with the experimental data, which allows us to conclude that the proposed model as a whole correctly describes both the annihilation process and the FSIs of annihilation-generated mesons; thus, it is expected that there should be small uncertainties associated with the simulation of an intranuclear $\bar{n}$\isotope[15][8]{O} annihilation.
    
    \item During the propagation of cascade particles (mesons and nucleons), the nucleus accumulates excitation energy. The final stage of $\bar{n}$A annihilation is thus the de-excitation of the residual nucleus. In Sec.~\ref{sec:SMM}, a new implementation describing de-excitations of residual nuclei with $\gamma$ emission is presented for the first time. A de-excitation model, taking into account gamma emission, does currently exists within MCSK.
\end{enumerate}

\begin{table}[ht!]
    \caption{Meson multiplicity comparisons for elementary $p\bar{p}$ annihilation using this generator~\citep{Golubeva:2018mrz,Barrow:2019viz} and available data sets~\protect\citep{Salvini:2004gz,Amsler:2003bq}.}
    \label{tab:newmesonmultiplicities}
    \centering
    \begin{ruledtabular}
    \begin{tabular}{c|cc}
         & $\bar{p}p$ Sim. & $\bar{p}p$ Exp. \\ 
         \hline
        $M(\pi)$ & $4.95$ & $4.98\pm0.35$~\protect\citep{Klempt:2005pp}, $4.94\pm0.14$~\protect\citep{Minor:1990mk}\\
        $M(\pi^{\pm})$ & $3.09$ & \footnotesize{$3.14\pm0.28$~\protect\citep{Klempt:2005pp}, $3.05\pm0.04$~\protect\citep{Klempt:2005pp}, $3.04\pm0.08$~\protect\citep{Minor:1990mk}}\\
        $M(\pi^0)$ & $1.86$ & \footnotesize{$1.83\pm0.21$~\protect\citep{Klempt:2005pp}, $1.93\pm0.12$~\protect\citep{Klempt:2005pp}, $1.90\pm0.12$~\protect\citep{Minor:1990mk}}\\
        $M(\eta)$ & $0.09$ & $0.10\pm0.09$~\protect\citep{Levman:1979gg}, $0.07\pm0.01$~\protect\citep{Klempt:2005pp}\\
        $M(\omega)$ & $0.27$ & $0.28\pm0.16$~\protect\citep{Levman:1979gg}, $0.22\pm0.01$~\protect\citep{Hamatsu:1976qz}\\ 
        $M(\rho^+)$ & $0.19$ & -----\\ 
        $M(\rho^-)$ & $0.18$ & -----\\ 
        $M(\rho^0)$ & $0.18$ & $0.26\pm0.01$~\protect\citep{Hamatsu:1976qz}
    \end{tabular}
    \end{ruledtabular}
\end{table}


\begin{table*}[ht!]
    \caption{A list of updated multiplicities from experimental data and the model for $\bar{p}$\isotope[12][6]{C}, taking into account all annihilation branching ratios, the intranuclear antinucleon potential, and an associated nuclear medium response~\citep{Golubeva:2018mrz,Barrow:2019viz}. Averages are shown for 100,000 events.}
    \label{tab:multiplicities-exper_calc}
    \centering
    \begin{ruledtabular}
    \begin{tabular}{c|ccccccc}
         & $M(\pi)$ & $M(\pi^+)$ & $M(\pi^-)$ & $M(\pi^0)$ & $E_{tot}$\,(MeV) & $M(p)$ & $M(n)$\\ 
         \hline
         $\bar{p}{\rm C}$ Experiment & $4.57 \pm 0.15$ & $1.25 \pm 0.06$ & $1.59 \pm 0.09$ & $1.73 \pm 0.10$ & $1758 \pm 59$ & ----- & -----\\
         $\bar{p}{\rm C}$ Calculation & $4.60$ & $1.22$ & $1.65$ & $1.73$ & $1762$ & $0.96$ & $1.03$\\
    \end{tabular}
\end{ruledtabular}
\end{table*}

\twocolumngrid

\section{Nuclear effects in \isotope[16][8]{O}} \label{sec:nuceffects} 
\subsection{Annihilation density for \isotope[16][8]{O}}
The formalism of intranuclear $n\rightarrow\bar{n}$ has been explained in several papers~\citep{Sandars:1980pr,Dover:1982wv,Friedman:2008es,Barrow:2019viz}. Thus, we restrict ourselves here to a brief reminder of key features of the physics at play. The model of the \isotope[16][8]{O} nucleus is described in terms of an effective and realistic shell model, tuned to describe its main properties. Each $n$ shell is characterized by its radial number, $n$, orbital angular momentum, $\ell$, and total angular momentum, $J$ (after suitable spin-orbit coupling), resulting in the radial equation
\begin{equation}\label{eq:neutron-shell}
\begin{gathered}
-\frac{u_i''(r)}{2\,\mu}+ \left[\frac{\ell(\ell+1)}{2\,\mu\,r^2}+U_i(r)-\epsilon_i\right]\,u_i(r)=0 \, , \\
u_i(0)=u_i(\infty)=0 \, ,
 \end{gathered}
\end{equation}
where $i$ collectively stands for the quantum numbers $\{n,\ell,J\}$, and $\mu$ is the reduced mass of the neutrons with respect to the rest of the nucleus. A small $\bar{n}$ component, $v_i(r)$, is attached to each $n$ shell, which, to the first non-vanishing order, is given by the inhomogoneous equation
\begin{equation}
\begin{gathered}
 -\frac{v_i''(r)}{2\,\mu}+ \left[\frac{\ell(\ell+1)}{2\,\mu\,r^2}+V(r)-\epsilon_i\right]\,v_i(r)=\gamma\,u_i(r) \, ,\\
 v_i(0)=v_i(\infty)=0 \, .
 \end{gathered}
\end{equation}
Once the $\bar{n}$ radial wave function $v_i$ is calculated, its contribution $\Gamma_i$ to the width is given by 
\begin{equation}\label{eq:antineutron-shell}
-\frac{\Gamma_i}{2}=\int\limits_0^\infty |v_i(r)|^2\,\Im V\,{\rm d}r
=\gamma\,\int\limits_0^\infty u_i(r)\,\Im v_i(r)\,{\rm d}r~.
\end{equation}
\noindent
The inputs considered here for the calculation of the radial annihilation probability distribution are the $n$ wave functions $u_i(r)$, given by the effective shell model~\citep{Bolsterli:1972zz,Ajzenberg-Selove:1977gcr}, the corresponding shell energies $\epsilon_i$, and the complex antineutron-nucleus potential $V$. The strength $\delta$ of the transition is $\delta=1/\tau_{n\bar{n}}$ (which is unknown), and the potential $V$ is determined by a fit to data on antiprotonic atoms and antiproton-nucleus scattering; for a review and references, see, e.g.,~\citep{Richard:2019dic}. 

In the present calculation, a simple form for the optical potential $V$ has been adopted, where
\begin{equation}\label{eq:our-opt-pot}
    V(r)=-\frac{4\pi}{\mu}\rho(r)\,b~,
\end{equation}
where $\rho(r)$ is the local nuclear density calculated by the radial functions $u_i(r)$ and their analogs for protons, and $b$ is an effective scattering length taken as $b=1.3+i\, 1.9\,$fm~\citep{Friedman:2008es}.

\begin{figure}[h!]
  \centering
  \includegraphics[width=.9\columnwidth]{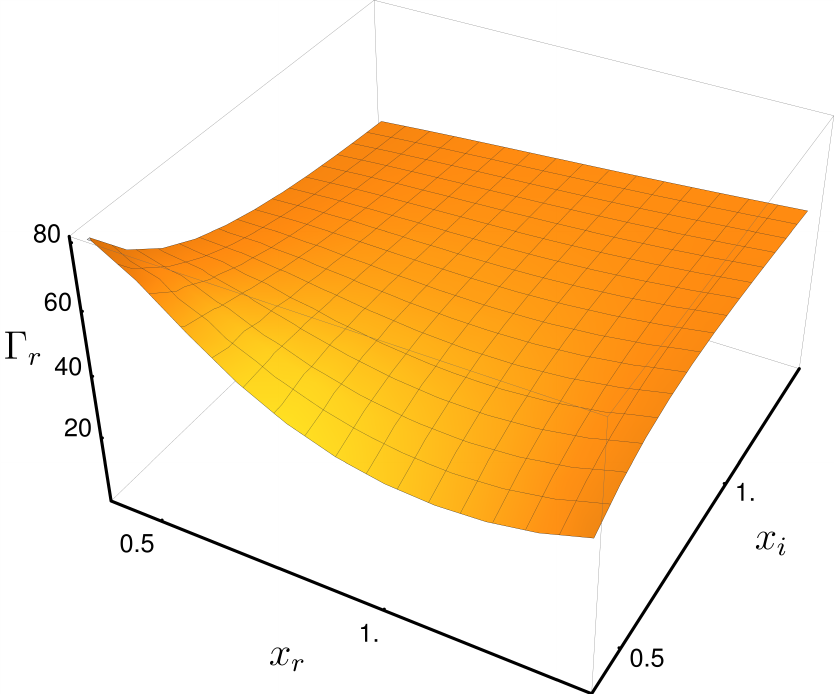}
  \caption{The change in the width $\Gamma_{r}$ for a simple model when the real part of the antineutron-nucleus potential is modified by a factor $x_{\rm r}$ and its imaginary part by $x_{rm i}$.}
  \label{fig:nnbar:stab}
\end{figure}

The most remarkable property of Eq.~\eqref{eq:antineutron-shell} is its stability with respect to changes in the $\bar{n}$ potential; Fig.~\ref{fig:nnbar:stab} shows how the width varies in a simple model when the real ($x_r$) and imaginary ($x_i$) parts of the $\bar{n}$ potential $V$ are varied, namely
$V \to [ {x_r}\,\Re(V)+ i\, {x_i}\,\Im(V) ]$, where $x_{i,r}$ are continuous factors representing a percentile change.

Another concern is the use of antiproton-nucleus data to determine the $\bar{n}$ interaction, $U$. The optical potential $V$ results from folding the antinucleon-nucleon ($\bar{N} N$) amplitude $\mathcal{M}$ with the distribution of the nucleons within the nucleus. However, it is known that $\mathcal{M}$ is weakly isospin dependent, as shown by the smallness of the charge-exchange cross-section $\bar{p} p \to \bar{n} n$ which is governed by $\mathcal{M}_{I=0}-\mathcal{M}_{I=1}$.

\begin{figure}[ht!]
    \centering
    \includegraphics[width=1.0\columnwidth]{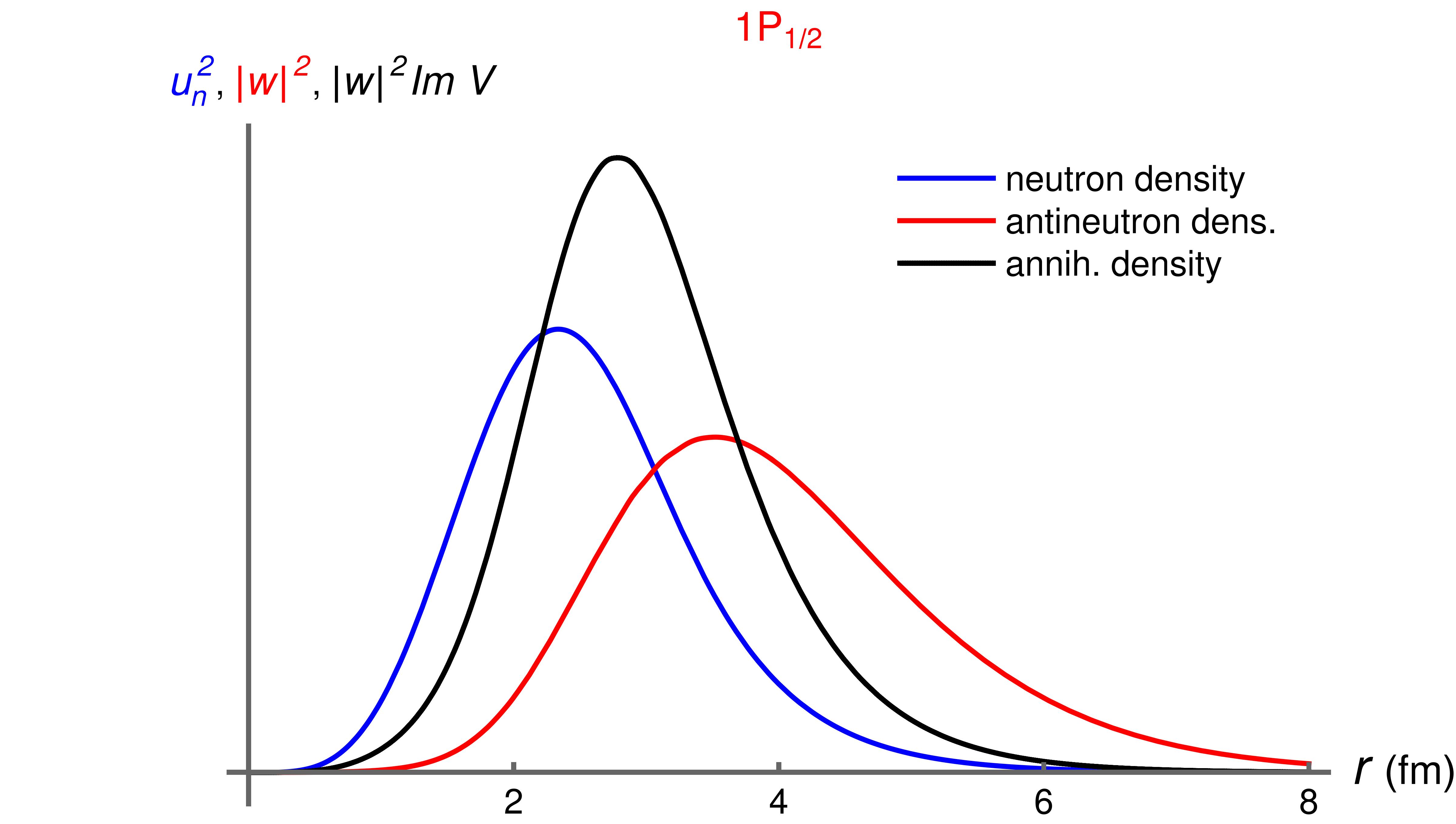}
    \caption{Radial distributions for the ${}^1\mathrm{P}_{1/2}$ shell of \isotope[16][8]{O}: $n$, $\bar{n}$, and annihilation. The units are arbitrary for the vertical axis. The full annihilation density shown in dashed black is the same as will be shown later in Fig.~\ref{fig:AnnRadProbDistsO}.}
    \label{fig:radial_dists}
\end{figure}

The most comforting observation is that the $n\rightarrow\bar{n}$ and the subsequent $\bar{n}N$ annihilation occurs at the surface of the nucleus, precisely the same region probed by antiprotonic atoms and antinucleon scattering experiments
. An illustration is given in Fig.~\ref{fig:radial_dists} for the case of the ${}^1 \mathrm{P}_{1/2}$ shell for \isotope[16][8]{O}. The $\bar{n}$, generated {\`a} la Eq.~\eqref{eq:antineutron-shell}, is clearly seen to be outside the $n$ distribution. However, the $\bar{n}$ requires some partner to annihilate with, and so as a compromise the annihilation takes place at the surface. 

Altogether, from these and past studies~\citep{Friedman:2008es,Barrow:2019viz}, the nuclear part of the calculation is viewed to be rather safe.

\subsection{Comparisons of intranuclear suppression factors for \isotope[16][8]{O}} \label{sec:suppression_comparison}
From the above formalism, one obtains a suppression factor (averaged over all neutrons) of $T_\text{R}=0.65\times10^{23}\,\text{s}^{-1}$, which is comparable with the estimates $0.65\times10^{23}\,\text{s}^{-1}$ discussed in~\citep{Friedman:2008es} and $0.8\times10^{23}\,\text{s}^{-1}$ in the early calculation by Dover et al.~\citep{Dover:1982wv}. Adopting the second model of that latter reference, 
\begin{equation}\label{eq:DoverII}
V=-\frac{U_\text{R}+i\,U_\text{I}}{1+\exp[(r-R)/a)]}~,
\end{equation}
with $U_\text{R}=107\,$MeV, $U_\text{I}=222\,$MeV, $R=2.388\,$fm and $a=0.523\,$fm, one arrives at exactly the same $T_\text{R}=0.65\times10^{23}\,\text{s}^{-1}$. This implies that the small discrepancy with respect to~\citep{Dover:1982wv} comes here from a choice of better $n$ wave functions.

Note that the effective scattering length used in~\citep{Friedman:2008es} has been adopted; however, in this work, a finite range is introduced for the folding with the nuclear density
. It has been checked that, indeed, increasing the size of the optical potential somewhat reduces the suppression factor $T_\text{R}$. Moreover, it has been checked that the antinucleon wave functions and the radial annihilation distribution, such as the one shown in Fig.~\ref{fig:radial_dists}, are very similar to the ones displayed by Friedman and Gal~\citep{Friedman:2008es}.

\section{Statistical description of nuclear disintegration} \label{sec:SMM}
Statistical approaches have proven very successful for describing the evolution of excited nuclei. According to the statistical hypothesis, any initial dynamical interactions between nucleons lead to a redistribution of the available energy among many degrees of freedom, and the nuclear system evolves towards equilibrium. The most famous example of such an equilibrated nuclear system is the `compound nucleus' introduced by Niels Bohr in 1936~\citep{Bohr:1936zz}. However, as was more recently established, the further evolution of thermal nuclear systems depend crucially on their excitation energy and mass number. The standard compound nucleus picture is valid only at low excitation energies and for large nuclei when sequential evaporation of light particles and fission are the dominant decay channels~\citep{Bondorf:1995ua,Botvina:2005aw,Eren:2013lda}. The concept of the compound nucleus cannot be applied at high excitation energies, $E^* \goo 3\,$MeV/nucleon, the reason being that time intervals between subsequent fragment emissions can become very short at $\sim10\,$fm/c $\sim3.3\times 10^{-23}\,$s. In this case there will not be enough time for the residual nucleus to reach equilibrium between subsequent emissions. Moreover, the nuclear fragments produced will be in the vicinity of each other, and, therefore, should interact together strongly. Thus, the rates of particle emission generally calculated for an isolated compound nucleus will not be reliable in this situation.

There are many other theoretical and experimental arguments in favour of a simultaneous break-up at high excitation energy for light excited nuclei~\citep{Bondorf:1995ua,Botvina:2005aw}. Since after an intranuclear $\bar{n}N$ annihilation one can expect a spectrum of excitation energies from low to very high energies, the de-excitation code used here it taken to be the Statistical Multifragmentation Model (SMM), mainly described in review~\citep{Bondorf:1995ua}. This model includes all de-excitation processes which can be present in such excited nuclei: the multifragmentation of larger nuclei at high energies, their evaporation and fission at low energies, and the simultaneous break-up of light excited nuclei (so called Fermi-break-up). All these processes are properly connected and self-consistent within the code. The code has demonstrated a very good description of various experimental data~\citep{Bondorf:1995ua,Botvina:1994vj,Scharenberg:2001gx,DAgostino:1995sny,Bellaize:2002cz,Avdeyev:2002fp,Pienkowski:2002qh,DAgostino:1999nlf}. 

At high excitation energies, the SMM assumes statistical equilibrium of the nuclear system with mass number $A_0$, charge $Z_0$, and excitation energy $E_0$ within a low-density freeze-out volume. This volume can be parameterized as $V=V_0+V_f$, so the baryon density is $\rho = A_0 / V$; here, $V_0$ is the volume of the system at a normal nuclear density $\rho_0 \approx 0.15\,$ fm$^{-3}$, and $V_f$ is the so-called free volume available for any translational motion of nuclear fragments. In the excluded volume approximation, $V_f$ may be taken as a constant for all break-up channels; however, under more realistic assumptions, it is known to depend on the nuclear fragment multiplicity, $M$, within each of the channels~\citep{Bondorf:1995ua}. The model considers all break-up channels (an ensemble of partitions $\{p\}$) composed of nucleons and excited fragments, all taking into account the conservation of baryon number, electric charge, and energy. An important advantage of the SMM is that besides its many multifragment break-up channels, it also includes compound nucleus decay channels, and accounts for the rate competition between all these channels. 

For the freeze-out volume, light nuclei with mass numbers $A \leq 4$ and charges $Z \leq 2$ are treated as elementary stable particles with masses and spins taken from the nuclear tables (a ``nuclear gas"). Only the translational degrees of freedom of these particles contribute to the entropy of the system. Fragments  with $A > 4$ are treated as heated nuclear liquid drops; in this way, one may study the coexistence of both nuclear liquid-gas phases in the freeze-out volume. Their individual free energies $F_{AZ}$ are parameterized as a sum of the bulk, surface, Coulomb, and symmetry energy contributions:
\begin{equation}
    F_{AZ}=F^{B}_{AZ}+F^{S}_{AZ}+E^{C}_{AZ}+E^{sym}_{AZ}\,.
\end{equation}
\noindent
The standard expressions for these terms are: $F^{B}_{AZ}=(-W_0-T^2/\epsilon_0)A$, where $T$ is the temperature, the parameter $\epsilon_0$ is related to the level density, and $W_0 = 16\,$MeV is the binding energy of infinite nuclear matter; $F^{S}_{AZ}=B_0A^{2/3}(T^2_c-T^2)^{5/4}(T^2_c+T^2)^{-5/4}$, where $B_0=18\,$MeV is the surface coefficient, and $T_c=18\,$MeV is the critical temperature of infinite nuclear matter; $E^{C}_{AZ}=cZ^2/A^{1/3}$, where $c=(3/5)(e^2/r_0)(1-(\rho/\rho_0)^{1/3})$ is the Coulomb parameter (obtained in the Wigner-Seitz approximation)~\citep{Bondorf:1995ua}, with the charge unit $e$ and $r_0=1.17\,$fm; $E^{sym}_{AZ}=\gamma(A-2Z)^2/A$, where $\gamma = 25\,$MeV is the symmetry energy parameter. These parameters are those of the Bethe-Weizs\"acker formula, corresponding to the assumption of isolated fragments with normal density in the freeze-out configuration; such an assumption has been found to be quite successful in many applications~\citep{Bondorf:1995ua}. It is to be expected, however, that in a more realistic treatment one must expand this description, whereby primary nuclear fragments will have to be considered not only excited but also subject to a nuclear interaction between them. These effects can be accounted for in the fragment free energies by changing the corresponding liquid-drop parameters. The Coulomb interaction of fragments in the freeze-out volume is described within the Wigner-Seitz approximation; see ref.~\citep{Bondorf:1995ua} for details. 

As is well known, the number of partitions of medium and heavy systems $(A_0\sim 100)$ is enormous~\citep{Bondorf:1995ua}. In order to take these into account, the SMM uses a few prescriptions. At small excitation energies, the standard SMM code~\citep{Bondorf:1995ua} uses a microcanonical treatment, though taking into account a limited number of disintegration channels: as a rule, only partitions with total fragment multiplicity $M \leq 3$ are considered. This is a very convenient approximation at low temperature, wherein the compound nucleus and low-multiplicity channels dominate, permitting a smooth transition into these channels. Recently, a full microcanonical version of the SMM using a Markov Chain Monte Carlo method was introduced~\citep{Botvina:2000jc}. Thus, it can be used for exploring all partitions without limitation.

Within the microcanonical ensemble, the statistical weight of a partition $p$ is calculated as 
\begin{eqnarray}
W_{\rm p} \propto e^{S_{\rm p}} \, ,
\end{eqnarray}
\noindent
where $S_{\rm p}$ is the corresponding entropy; this further depends on fragments in this partition, as well as on the excitation energy $E_0$, mass number $A_{0}$, charge $Z_{0}$, and volume $V$ of the residual nuclear system. In the standard treatment, a description which corresponds to an approximate microcanonical ensemble is followed; namely, a temperature $T_{p}$ characterising all final states in each partition $p$ is introduced. This is determined from the energy balance equation taking into account the total excitation energy $E_0$~\citep{Bondorf:1995ua}. In the following, $S_{\rm p}$ for the calculated $T_{p}$ is determined by using conventional thermodynamical relations~\citep{Bondorf:1995ua}. In the standard case, it can be written as 
\begin{equation}
\begin{aligned}
S_{\rm p} = & \ln{\left(\prod_{A,Z}g_{A,Z}\right)} + \ln{\left(\prod_{A,Z}A^{3/2}\right)} - \ln{\left(A_0^{3/2}\right)} \\
            & - \ln{\left(\prod_{A,Z}n_{A,Z}!\right)} + (M-1)\ln{\left(\frac{V_f}{\lambda_{T_{p}}^3}\right)} \\
            & + 1.5(M-1)+\sum_{A,Z}\left(\frac{2T_{p}A}{\epsilon_0} - \frac{\partial F^{S}_{AZ}(T_{p})}{\partial T_{p}}\right) \, ,
\end{aligned}
\end{equation}
\noindent
where $n_{A,Z}$ as the number of fragments with mass $A$ and charge $Z$ in the partition, $g_{A,Z}=(2s_{A,Z}+1)$ is the spin degeneracy factor, $\lambda_{T_{p}}=\left(2\pi\hbar^2/m_NT_{p}\right)^{1/2}$ is the nucleon thermal wavelength ($m_N\approx 939$ MeV is the average nucleon mass), and the summation is performed over all fragments of the partition $p$. One may enumerate all considered partitions and select one of them according to its statistical weight by the Monte Carlo method.

At very high excitation energy, the standard SMM code makes a transition to the grand-canonical ensemble~\citep{Bondorf:1995ua} since the number of partitions with high probability becomes too large. In the grand canonical formulation, after integrating out translational degrees of freedom, one can write the mean multiplicity of nuclear fragments with $A$ and $Z$ as
\begin{eqnarray}
\label{naz} \langle n_{A,Z} \rangle =
\frac{g_{A,Z} V_{f} A^{3/2}}{\lambda_T^3} \, e^{-\left(F_{AZ}(T,V)-\mu A-\nu Z\right)/T}\, . 
\end{eqnarray}
\noindent
Here, the temperature $T$ can be found from the total energy balance of the system by taking into account all possible fragments with $A$ from $1$ to $A_0$, and with $Z$ from 0 to $Z_0$~\citep{Bondorf:1995ua}. The chemical potentials $\mu$ and $\nu$ are found from the mass and charge constraints:
\begin{equation} \label{eq:ma2}
\sum_{A,Z}\langle n_{A,Z}\rangle A=A_{0}, \, \, \, 
\sum_{A,Z}\langle n_{A,Z}\rangle Z=Z_{0} \, .
\end{equation}
\noindent
In this case, the grand canonical occupations $\langle n_{A,Z} \rangle$ are used for Monte-Carlo sampling of the fragment partitions~\citep{Bondorf:1995ua}. These two methods of partition generation are carefully adjusted to provide a proper transition from the low energy to the high energy regimes~\citep{Bondorf:1995ua}.

After the Monte-Carlo generation of a partition, the temperature, excitation energy, and momenta can be found for ``hot" fragments from the energy balance equation
. In this approach, the temperature may slightly fluctuate from partition to partition since the total energy of the system $E_0$ is always conserved. At the next stage, Coulomb acceleration and propagation of the nuclear fragments must be taken into account. For this purpose, the fragments are placed randomly across the freeze-out volume $V$ (without overlapping), and their positions are adjusted by taking into account that their Coulomb interaction energy, which must be equal to the value calculated in the Wigner-Seitz approximation~\citep{Bondorf:1995ua}. Note that, in the case of the Markov Chain SMM version~\citep{Botvina:2000jc}, this adjustment is not necessary since the positions of the nuclear fragments are sampled directly. Subsequently, the method employed here re-solves the Hamilton equations of motion for all of the nuclear fragments from these initial positions in their mutual Coulomb field. At this stage, a possible collective flow of nuclear fragments can be also taken into account~\citep{Bondorf:1995ua}. Usually this is done by adding supplementary radial velocities to the fragments (proportional to their distances from the centre of mass) at the beginning of Coulomb acceleration. The energy and momentum balances are strictly respected during this dynamical propagation. 

The secondary de-excitation of the primary hot fragments (including the compound nucleus) involves several mechanisms. For light primary nuclear fragments (with $A\leq 16$), even a relatively small excitation energy can be comparable to the fragments' total binding energies. In this case, it is assumed that the principal mechanism of de-excitation is an explosive decay of the excited nucleus into several smaller nuclei (the Fermi break-up)~\citep{Botvina:1987jp,Bondorf:1995ua}. In this decay, the statistical weight of the channel $p$ containing $n$  particles  with  masses $m_{i}$ (where $i=1,\cdots,n$) in a volume $V_{p}$ can be calculated in the microcanonical approximation as:
\begin{equation}
\begin{aligned}
\Delta\Gamma_{p}\propto  \frac{S}{G}& \left(\frac{V_{p}}{(2\pi\hbar)^{3}}\right)^{n-1}
\left(\frac{\prod_{i=1}^{n}m_{i}}{m_{0}}\right)^{3/2}
\\ &\times 
\frac{(2\pi)^
{\frac{3}{2}(n-1)}}{\Gamma(\frac{3}{2}(n-1))}  \left(E_{kin}-U_{p}^{C}\right)^{\frac{3}{2}n-\frac{5}{2}} \, ,
\end{aligned}
\label{eq:Fer}
\end{equation}
\noindent
where $m_{0}=\sum_{i=1}^{n}m_{i}$ is the mass of the decaying nucleus, $S=\prod_{i=1}^{n}(2s_{i}+1)$ is the degeneracy factor ($s_{i}$ is the $i$-th particle spin), $G=\prod_{j=1}^{k}n_{j}!$ is the particle identity factor ($n_{j}$ is the number of particles of kind $j$), $E_{kin}$ is the total kinetic energy of  particles at infinity (which can be found through the energy balance by taking into account the fragment excitation energy), and $U_{p}^{C}$ is the Coulomb barrier for the decay. Slight modifications to this model have been made by including nuclear fragment excited states which may remain stable with respect to any nucleon emission but which instead can decay via $\gamma$-emission afterwards. Also, some long-lived unstable nuclei (like $^5$He, $^5$Li, $^8$Be, $^9$B) are included, which decay later on. 
 
The successive particle emission from the primary hot fragments with $A>16$ is assumed to be the fundamental de-excitation mechanism, as in the case of a compound nucleus' decay. Due to the high excitation energy of these fragments, the standard Weisskopf evaporation scheme~\citep{Bondorf:1995ua} was modified to take into account any heavier ejectiles up to $^{18}$O (besides light particles such as nucleons, $d$, $t$, $\alpha$) in ground and particle-stable excited states~\citep{Botvina:1987jp}. The decay width for the emission of a particle $j$ from the compound nucleus $(A,Z)$ is given by:

\begin{equation}
\begin{aligned}
\Gamma_{j} = & \sum_{i=1}^{n}\int_{0}^{E_{AZ}^{*}-B_{j}-\epsilon_{j}^{(i)}} \frac{\mu_{j}g_{j}^{(i)}}{\pi^{2}\hbar^{3}}\sigma_{j}(E) \\
             & \times\frac{\rho_{A^{'}Z^{'}}(E_{AZ}^{*}-B_{j}-E)}{\rho_{AZ}(E_{AZ}^{*})} E dE \, .
\end{aligned}
\label{eq:eva}
\end{equation}
\noindent
Here, the sum is taken over all ground and particle-stable excited states $\epsilon_{j}^{(i)}~(i=0,1,\cdots,n)$ of the nuclear fragment $j$, $g_{j}^{(i)}=(2s_{j}^{(i)}+1)$ is the spin degeneracy factor of the $i$-th excited state, $\mu_{j}$ and $B_{j}$ correspond to the reduced mass and separation energy, $E_{AZ}^{*}$ is the excitation energy of the initial nucleus, $E$ is the kinetic energy of an emitted particle in the centre-of-mass frame, $\rho_{AZ}$ and $\rho_{A^{'}Z^{'}}$ are the level densities of the initial $(A,Z)$ and final $(A^{'},Z^{'})$ compound remnant nuclei, and $\sigma_{j}(E)$ is the cross section of the inverse reaction $(A^{'},Z^{'})+j=(A,Z)$ calculated using the optical model with a nucleus-nucleus potential~\citep{Botvina:1987jp}. The evaporation process was simulated via the Markov Chain Monte Carlo, method and the conservation of energy and momentum is strictly controlled in each emission step. 

At very low excitations when a nucleon emission is not possible, emission of photons is permitted. The decay width for the evaporation of $\gamma$-quanta from the excited remnant nuclei is taken in the statistical approximation as: 
\begin{equation} \label{eq:gam}
\Gamma_{\gamma}=\int_{0}^{E_{AZ}^{*}}
\frac{E^{2}}{\pi^{2}c^{2}\hbar^{2}}\sigma_{\gamma}(E)
\frac{\rho_{AZ}(E_{AZ}^{*}-E)}{\rho_{AZ}(E_{AZ}^{*})}dE \, .
\end{equation}
\noindent
This integration is performed numerically, and a dipole approximation for the photo-absorption cross section is used:
\begin{equation} \label{eq:gamsec}
\sigma_{\gamma}(E)=\frac{\sigma_{0}E^{2}\Gamma^{2}_{R}}
{(E^{2}-E^{2}_{R})^{2}+\Gamma^{2}_{R}E^{2}} \, . 
\end{equation}
\noindent
Here, $E$ is the $\gamma$ energy, and the empirical parameters of the giant dipole resonance take values $\sigma_0=2.5A\,$mb, $\Gamma_{R}=0.3E_{R}$, and $E_{R}=40.3/A^{0.2}\,$MeV. 

An important channel for the de-excitation of heavy nuclei ($A>100$) is fission. This process competes with particle emission, and it is also simulated within this Monte-Carlo method. Following the Bohr-Wheeler statistical approach~\citep{Bondorf:1995ua}, the partial width for a compound nucleus' fission is assumed to be proportional to the level density at the saddle point $\rho_{sp}(E)$:
\begin{equation} \label{eq:fis}
\Gamma_{f}=
\frac{1}{2\pi\rho_{AZ}(E_{AZ}^{*})}\int\limits_{0}^{E_{AZ}^{*}-B_{f}}
\rho_{sp}(E_{AZ}^{*}-B_{f}-E)dE\, ,
\end{equation}
\noindent
where $B_{f}$ is the height of the fission barrier determined from the Myers-Swiatecki prescription~\citep{Bondorf:1995ua}. For an approximation of $\rho_{sp}$, the results of an extensive analysis of nuclear fissility and $\Gamma_{n}$/$\Gamma_{f}$ branching ratios have been used~\citep{Bondorf:1995ua,Botvina:1994vj,Scharenberg:2001gx,DAgostino:1995sny,Bellaize:2002cz,Avdeyev:2002fp,Pienkowski:2002qh,DAgostino:1999nlf}. Other important details of the code one can find in Refs.~\citep{Bondorf:1995ua,Eren:2013lda}. 

Before connecting the SMM into simulations of $\bar{n}N$ annihilation, all these models for de-excitation were rigorously tested by numerical comparisons to experimental data on decays of compound nuclei with excitation energies less than $2$-$3\,$MeV per nucleon. It is important that after all stages the SMM provides event-by-event simulation for the whole break-up process, and allows for direct comparison with experimental events. Furthermore, like the parent intranuclear $\bar{n}N$ simulation, this code is modular and generic, allowing for many future isotopes to be studied.

\begin{figure}[ht!]
    \centering
    \includegraphics[width=1.0\columnwidth]{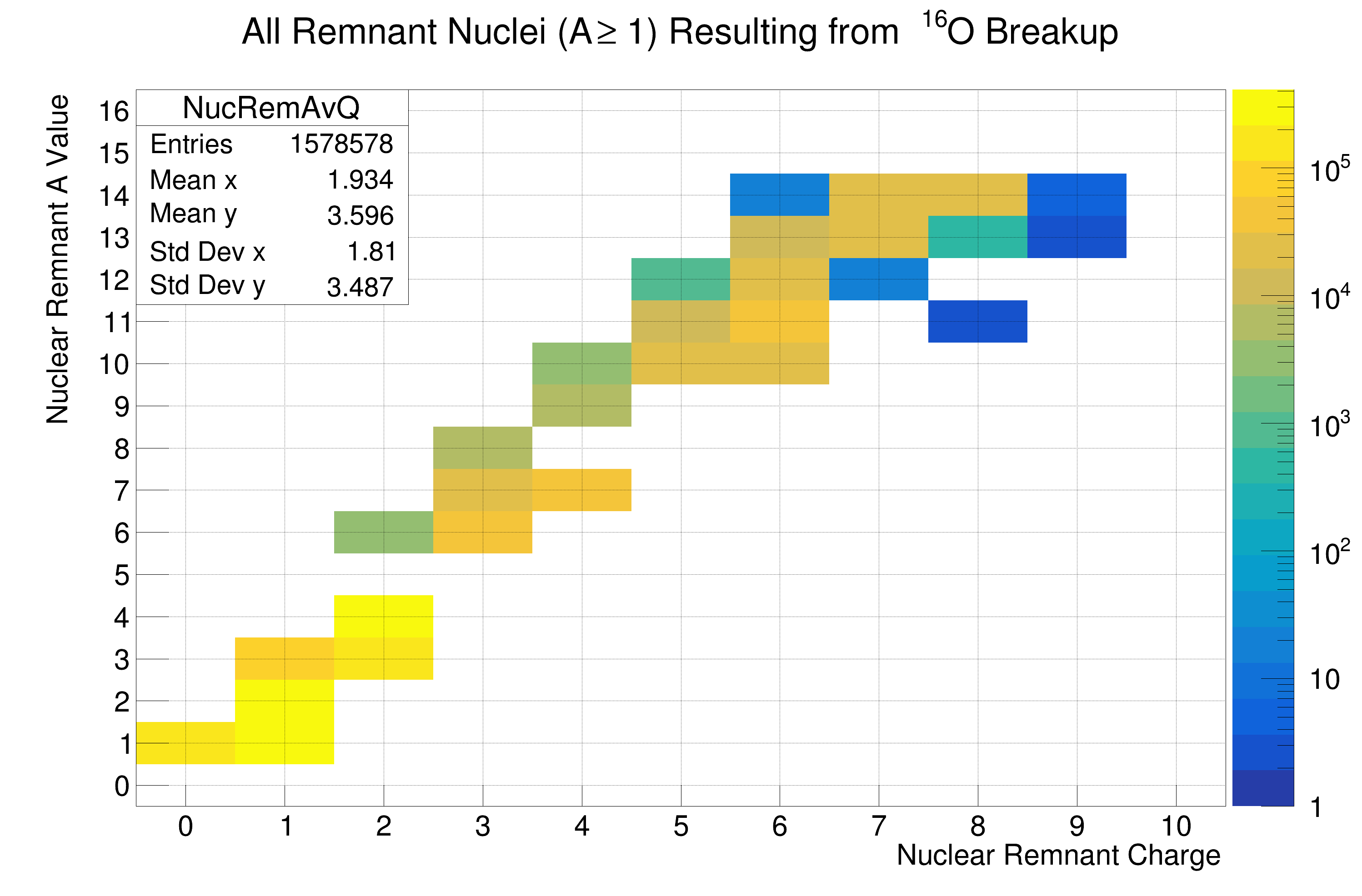}
    \caption{
    A correlation plot showing all intranuclear $n\rightarrow\bar{n}$ derived remnant nuclei with $A\geq 1$ following the breakup of the \isotope[16][8]{O} nucleus. On average, when ignoring evaporative particles with $A<2$, there are roughly two residual nuclei per annihilation event. Lack of $A=5$ nuclei is due to fast decays via strong interactions, and so do not produce photons; examples include $^5$He$\rightarrow ^4$He$+n$, $^5$Li$\rightarrow ^4$He$+p$, and $^8$Be$\rightarrow ^4$He$+ ^4$He. Shown for 500,000 generated events; note that some statistical fluctuations occur and populate rare states, for instance at $(A,Z)=\{(11,8),(13,9)\}$.}
    \label{fig:AllNucRemAvQ}
\end{figure}

\begin{figure}[h]
    \centering
    \includegraphics[width=1.0\columnwidth]{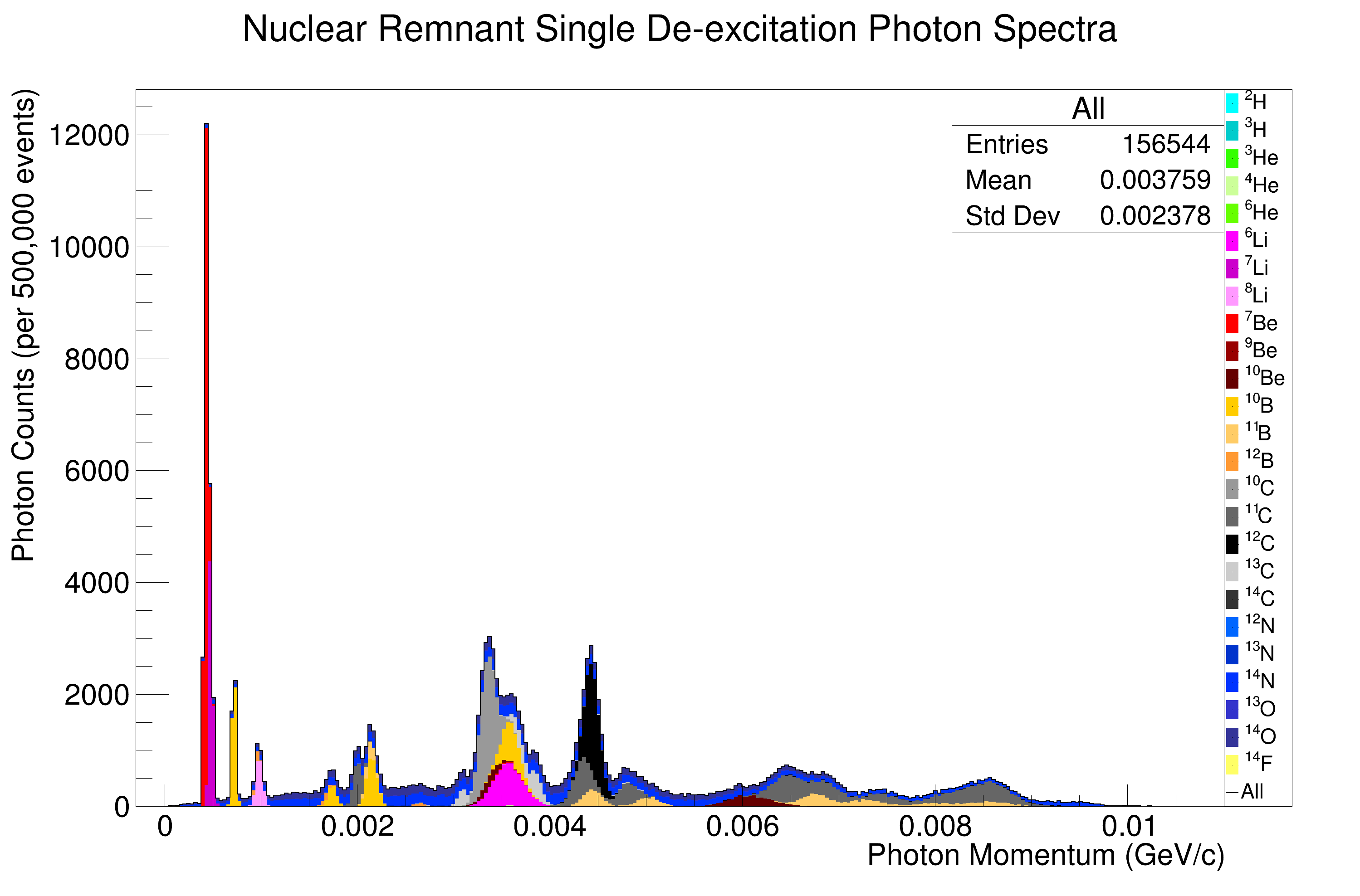}
    \includegraphics[width=1.0\columnwidth]{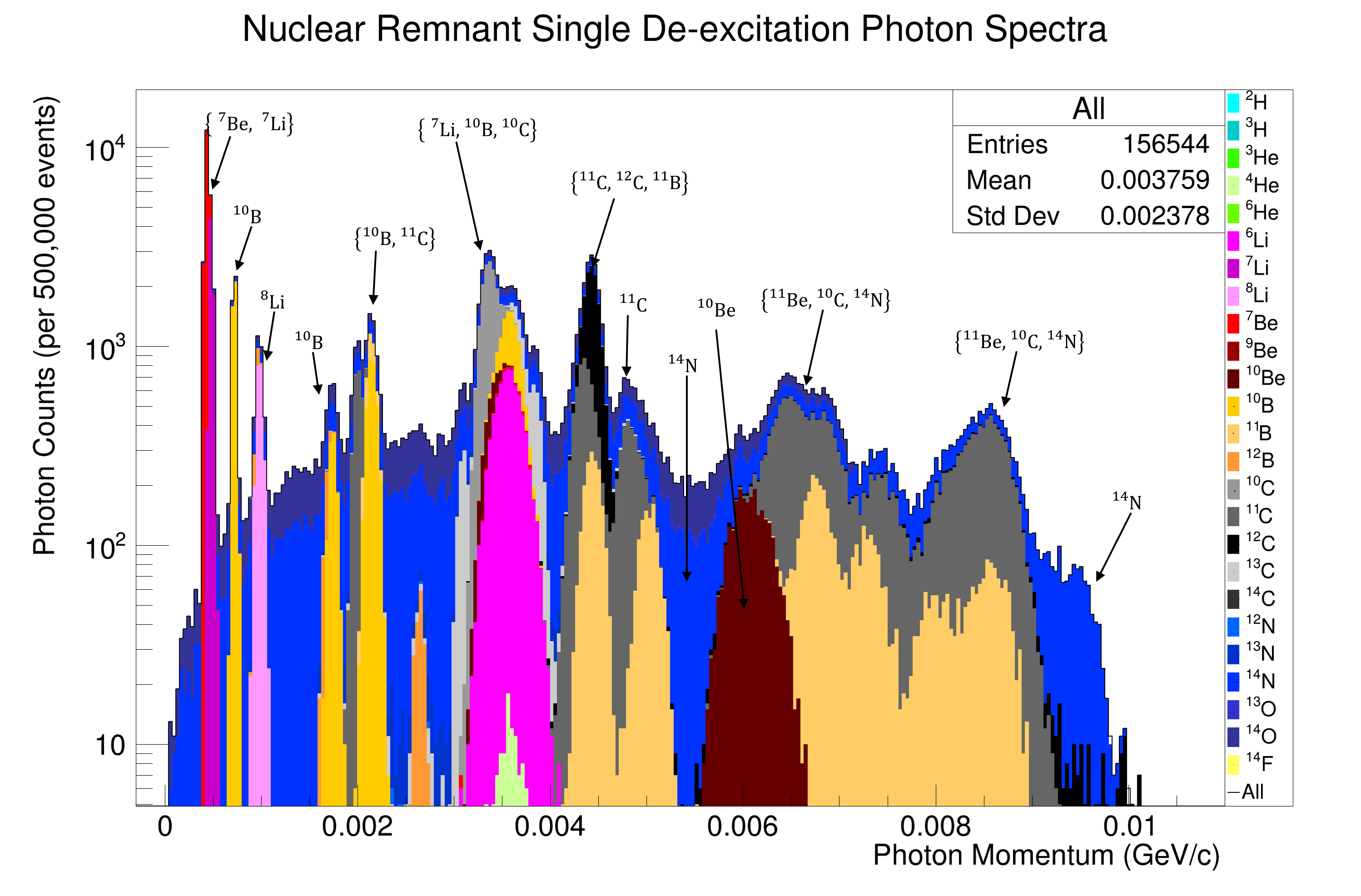}
    \caption{The single emission de-excitation photon spectrum of remnant nuclei (with $A \geq 2$) arising from the nuclear decay and evaporative processes following intranuclear $n\rightarrow\bar{n}$ in \isotope[16][8]{O}, shown in linear and logarithmic scales (labeled with some predominant nuclear isotopes for clarity). There do exist simulated events where $\geq 2$ photons are emitted, though these are produced exceedingly rarely and are not included here for simplicity. Shown for 500,000 generated events.}
    \label{fig:NucRemSingPhot}
\end{figure}

The abundant production of many nuclei in ground and excited states can be used for potentially improved identification of an intranuclear $n\rightarrow\bar{n}$ event; some of these can be seen in Fig~\ref{fig:AllNucRemAvQ}. The subsequent decays of nuclear remnants provide photons with different energies and different decay times, and can be in principle recognized with the advanced experimental methods. As an example, some population within the the high, narrow peaks shown in Fig.~\ref{fig:NucRemSingPhot} at $\gamma$ energies of $0.43\,$MeV and $0.47\,$MeV correspond to the first excited levels of $^7$Be and $^7$Li nuclei, each expected with the very short decay times of $133\,$fs and $77\,$fs, respectively\footnote{Note as well that, as seen from the chart of Fig.~\ref{fig:AllNucRemAvQ}, many nuclei in the ground state can decay as well; for example, ${}^{11}$C undergoes a positron emission decay, emitting a positron, and travels a short distance before colliding with an electron within the detector medium, producing a pair of gamma rays simultaneously in nearly opposite directions with an energy of $511\,$keV each. This half-life is around $20$ minutes. Another nucleus in the ground state, $^7$Be, will decay via the electron capture with a half-life of around $53$ days. All these processes provide an additional possibility for studying the phenomenon. These time scales are not necessarily useful for large underground liquid detectors due to fluid circulation and purification, but similar predictions for other nuclei could be useful within solid-state detectors as a confirmation of future activity at a potential vertex position of a candidate event.}.

\section{Generator validation} \label{sec:validation}
\begin{figure}
    \centering
    \includegraphics[width=1.0\columnwidth]{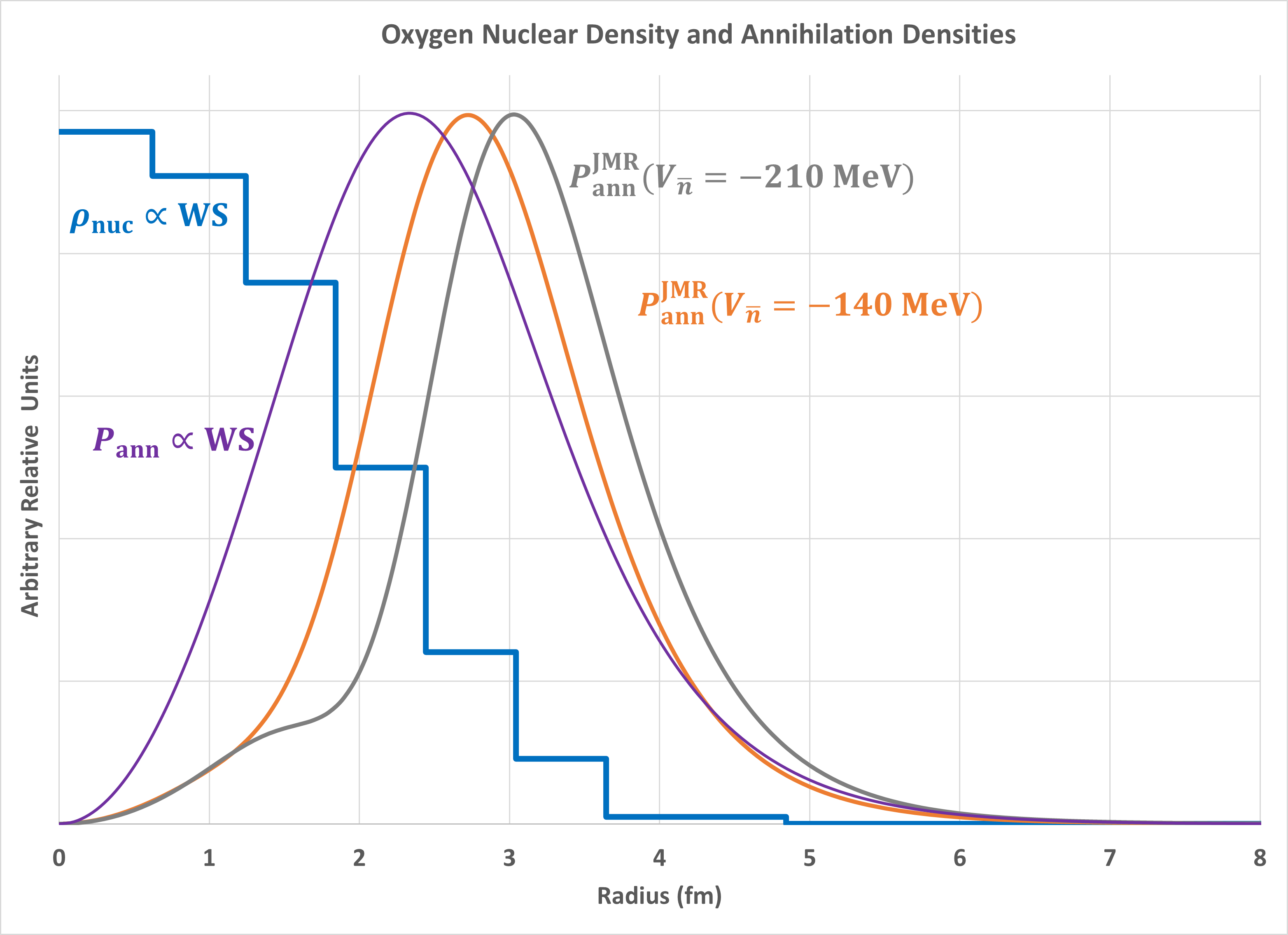}
    \caption{The radial annihilation probability distribution is shown for 
    \isotope[16][8]{O} using several models of the $\bar{n}$ potential well depth, including $V_{\bar{n}}(r=0)=-140\,$MeV (orange), $V_{\bar{n}}=-210\,$MeV (grey), and an annihilation distribution derived from the nuclear density (purple). Each of these are compared to the fitted, eight-zoned nuclear density distribution of \isotope[16][8]{O} in solid blue; see~\citep{Barrow:2019viz} for a discussion. All plots use an arbitrary vertical axis.}
    \label{fig:AnnRadProbDistsO}
\end{figure}

Here, some main features of the intranuclear $\bar{n}$\isotope[15][8]{O} annihilation simulation are discussed and presented. Consider again Fig.~\ref{fig:radial_dists} and its new cousin Fig.~\ref{fig:AnnRadProbDistsO}, where the radial distribution of nuclear density and of annihilation probability for different variants of the simulation are presented. It is seen that the radial annihilation probability is different for various generator assumptions and is dependent upon the value (depth) of $\bar{n}$ potential~\citep{Barrow:2019viz,Barrow:2021odz}, a free parameter of the model. Fig.~\ref{fig:AnnRadProbDistsO} shows a solid blue histogram of the radial distribution of the relative nuclear density for the oxygen nucleus used in the model alongside three variants of radial annihilation probability distributions in arbitrary units. In orange, a modern quantum mechanical formalism~\citep{Barrow:2019viz} is employed as discussed in Sec.~\ref{sec:nuceffects}, and the distribution is calculated for a well depth value of $V_{\bar{n}}(r=0)=-140\,$MeV for the attractive antinucleon potential. In purple a curve is shown simulating the same value of $V_{\bar{n}}(r=0)=-140\,$MeV, but the annihilation probability follows directly from the nuclear density, i.e. the quantum mechanical dynamics are not taken into account; without the antinucleon potential, this is somewhat similar to MCSK. In grey, the distribution shown is calculated for a value of $V_{\bar{n}}(0)=-210\,$MeV using the same modern formalism. It follows from these curves that the annihilation preferentially occurs within the diffuse peripheral layers of the nucleus, largely independent of these parameters. Importantly, though, it is seen  that a higher $\bar{n}$ potential shifts the annihilation distribution further outside the nucleus. This is somewhat counter-intuitive, but has an interesting explanation: if one has a weakly attractive potential, increasing its strength will gradually pull the wave function of a given particle inside. However, if one maintains a very strong absorptive potential, it repels the wave function. This is obvious for a strong repulsion; however, for a strong attraction, it creates nodes in the low-energy solution, and is effectively equivalent to a repulsion. Thus, this can actually force the $\bar{n}$ further \textit{out} of the nucleus. This point is a critical one, especially in understanding how annihilation generated pions might avoid final state interactions.

\begin{figure}
    \centering
    \includegraphics[width=1.0\columnwidth]{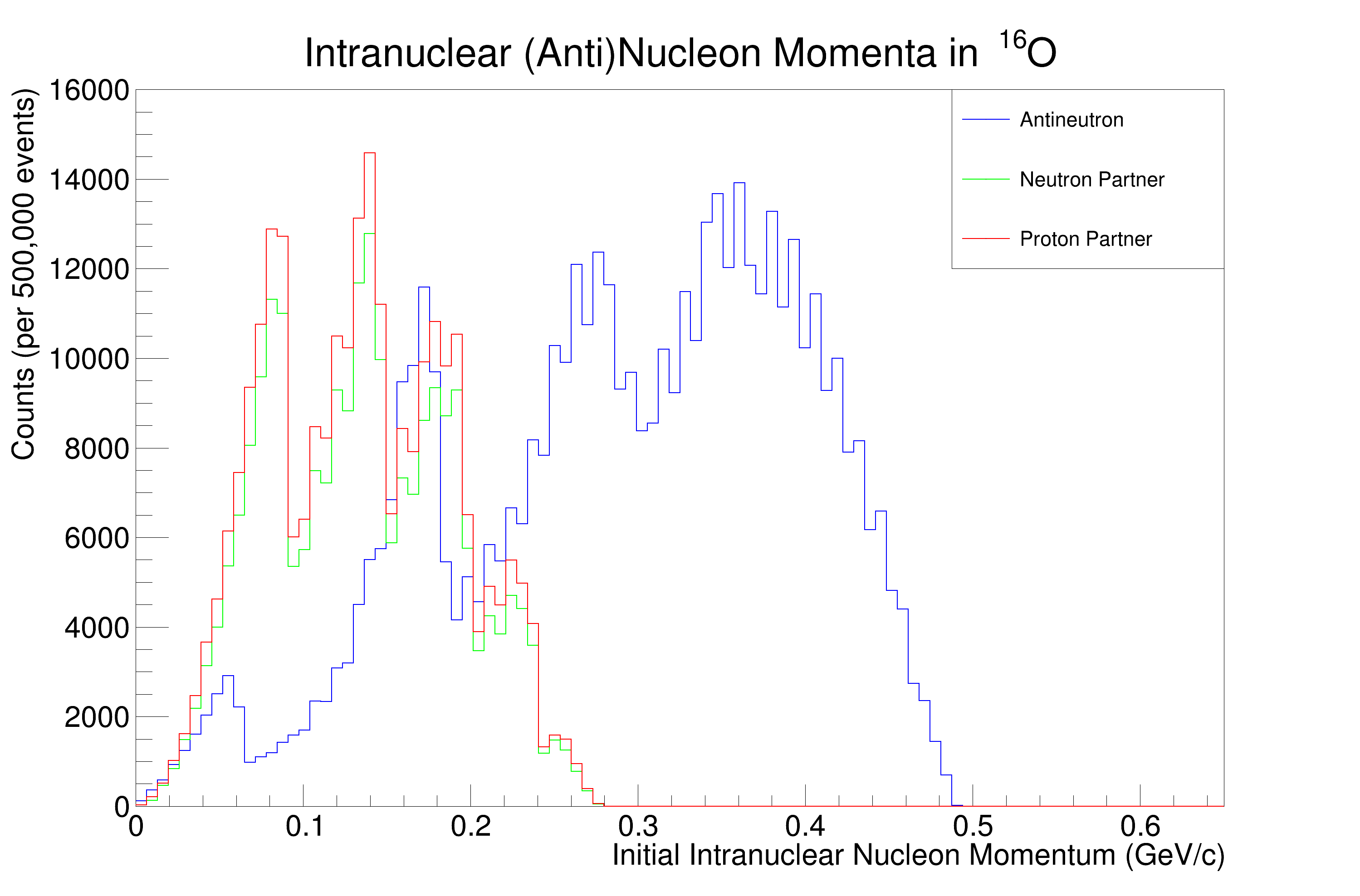}
    \caption{The initial momenta of intranuclear (anti)nucleons are shown for 500,000 generated events for an assumed $\bar{n}$ potential of $V_{\bar{n}}(r=0)=-140\,$MeV (orange in Fig.~\ref{fig:AnnRadProbDistsO}). $V_{\bar{n}}$ affects the annihilation probability distribution (Fig.~\ref{fig:AnnRadProbDistsO}) and is clearly visible from the larger Fermi momentum of the $\bar{n}$. The peaks throughout each nucleon momentum distribution are caused by the concentrically zoned structure of the nucleus, each with their own Fermi momentum; this structure becomes somewhat smeared out in the $\bar{n}$ case due to the presence of the extra antinucleon potential.}
    \label{fig:InitNucMom}
\end{figure}
\begin{figure}
    \centering
    \includegraphics[width=1.0\columnwidth]{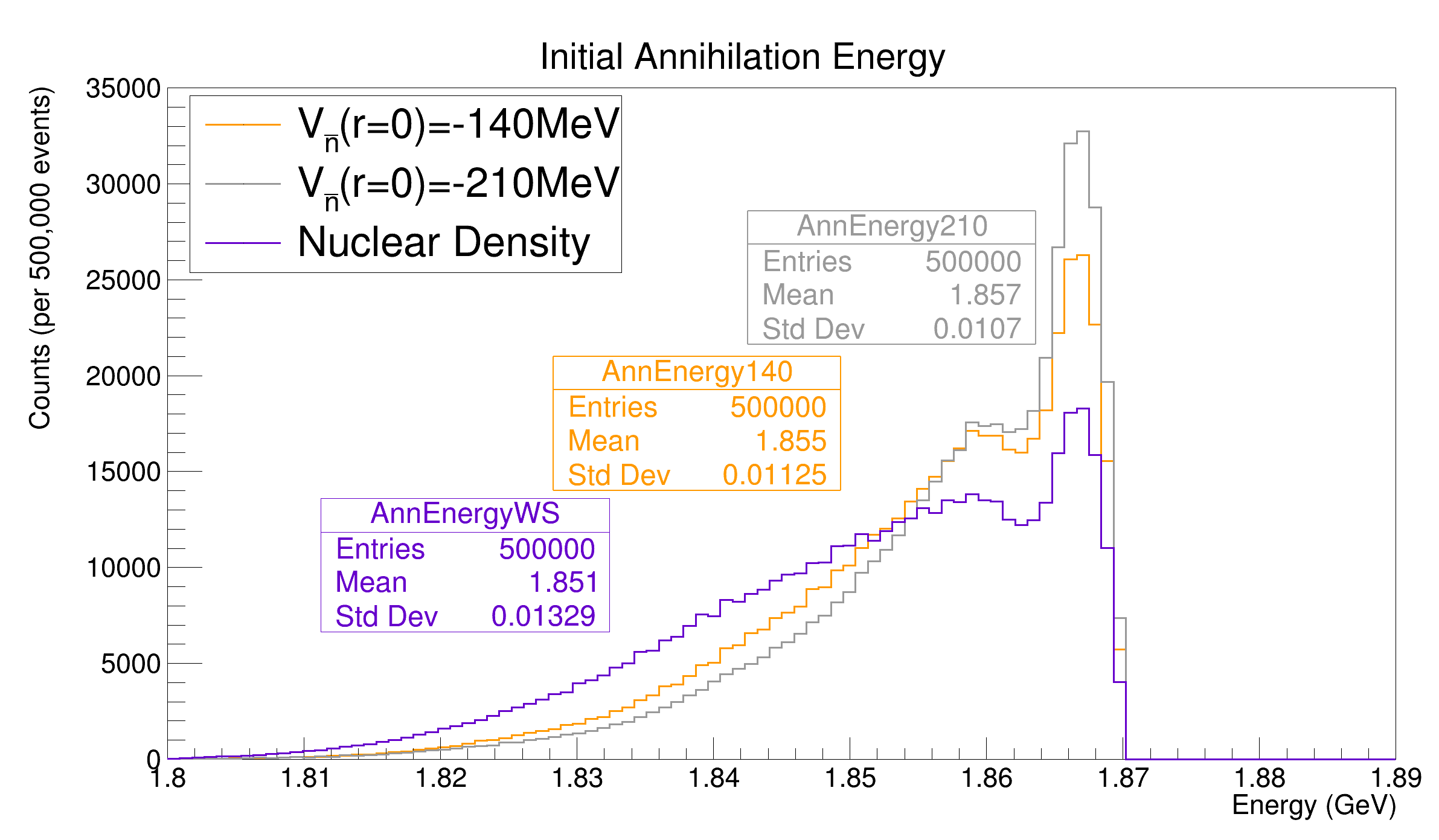}
    \caption{The initial total annihilation energy 
    of the $\bar{n}N$ pair and resulting annihilation-generated mesons within the nucleus are shown for several models of the $\bar{n}$ potential~\citep{Barrow:2019viz} and associated radial annihilation probability distribution (see again Fig.~\ref{fig:AnnRadProbDistsO}), including the modern annihilation densities with $V_{\bar{n}}(r=0)=-140\,$MeV (orange), $V_{\bar{n}}=-210\,$MeV (grey), and an annihilation distribution derived from the nuclear density (purple); note that the model derived from the nuclear density also assumes $V_{\bar{n}}(r=0)=-140\,$MeV. Each is shown for 500,000 generated events.}
    \label{fig:InitAnnE}
\end{figure}

With this in mind, now consider the histograms of Fig.~\ref{fig:InitNucMom}. Within this model of the nucleus, the momentum distribution of the nucleons within individual concentric nuclear zones are the same as for a degenerate Fermi gas; thus, the probability of a nucleon taking on a momentum $p$ in the $i$-th zone depends from it's boundary Fermi momentum, which in turn depends upon the local nuclear density. The nucleons located in the central zone of the nucleus have the maximum value of boundary Fermi momenta, $p_{FN}$, and generate the high-momentum part of the spectrum. Conversely, the nucleons located within the peripheral zone of the nucleus take on momenta of $\sim40$–$100\,$MeV/c; the most diffuse zone (beyond the periphery) occupies even lower initial values~\citep{Golubeva:2018mrz}. Of course, the larger the annihilation probability within any particular concentric zone, the greater the contribution to the final state pions' momentum spectra from the (anti)nucleons participating in the annihilation. Thus, in this model, there is a local correlation of the momentum with nuclear density, and, respectively, with the radius (for details, see~\citep{Golubeva:2018mrz}). The green and red histograms of Fig.~\ref{fig:InitNucMom} show the momentum distributions of the nucleons which are the annihilation partners within \isotope[15][8]{O}. The initial intranuclear momentum distribution for the $\bar{n}$ species following its transition (yet before the annihilation) is also presented in blue in Fig.~\ref{fig:InitNucMom}. Because of the introduction of an antinucleon potential and an associated off-shell mass for the $\bar{n}$, this distribution extends to higher momenta of $\sim 500\,$MeV/c~\citep{Barrow:2019viz}. Note that the histograms plotted here are considered for the annihilation density distribution derived from a potential well depth of $V_{\bar{n}}(r=0)=-140$, as seen in the orange curve of Fig.~\ref{fig:AnnRadProbDistsO}. Now consider the total energy available to the annihilation as presented in Fig.~\ref{fig:InitAnnE}. The plot shows that the total annihilation energy limit is independent of the depth of the antinucleon potential used and are indeed identical, as it must be from the energy conservation law $E_{\bar{n}}=E_{n}$ for any potential well depth following the $n\rightarrow\bar{n}$ transition. The average values of the total annihilation energy are close for all calculation options (from $1.851$-$1.857\,$GeV), however, the shape of the distributions differ from one assumption for the antinucleon potential to another, due to how the $\bar{n}N$ annihilation pair populations differ across various the nuclear radius (see again Fig.~\ref{fig:AnnRadProbDistsO}). From this, it can be interpreted that the more attractive the potential and peripheral the annihilation process is, the larger the proportion of events with high energy values, independent of final state interactions.

\newpage

\onecolumngrid

\begin{figure}
    \centering
    \includegraphics[width=0.49\columnwidth]{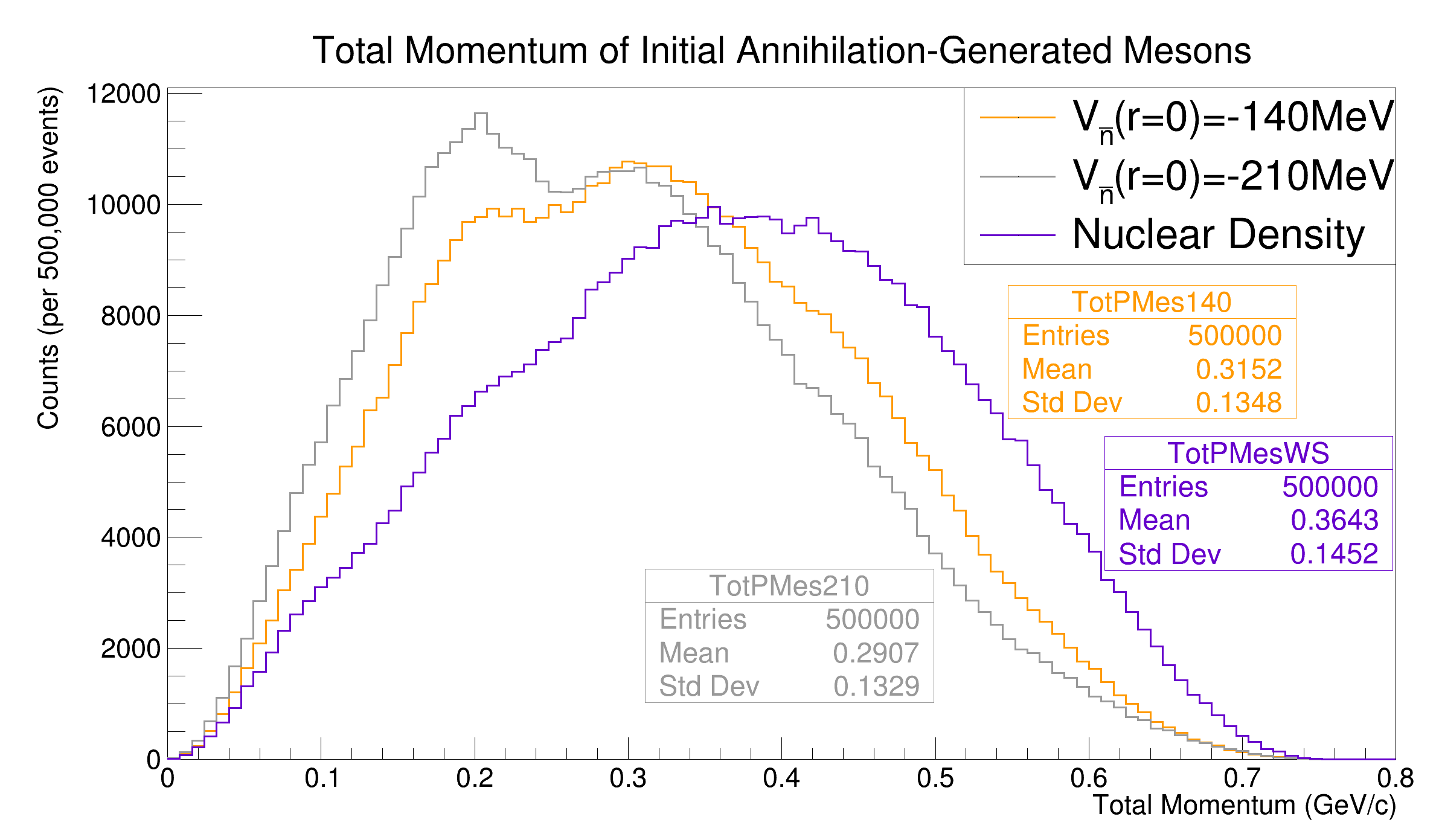}
    \hspace{0.04\columnwidth}
    \includegraphics[width=0.45\columnwidth]{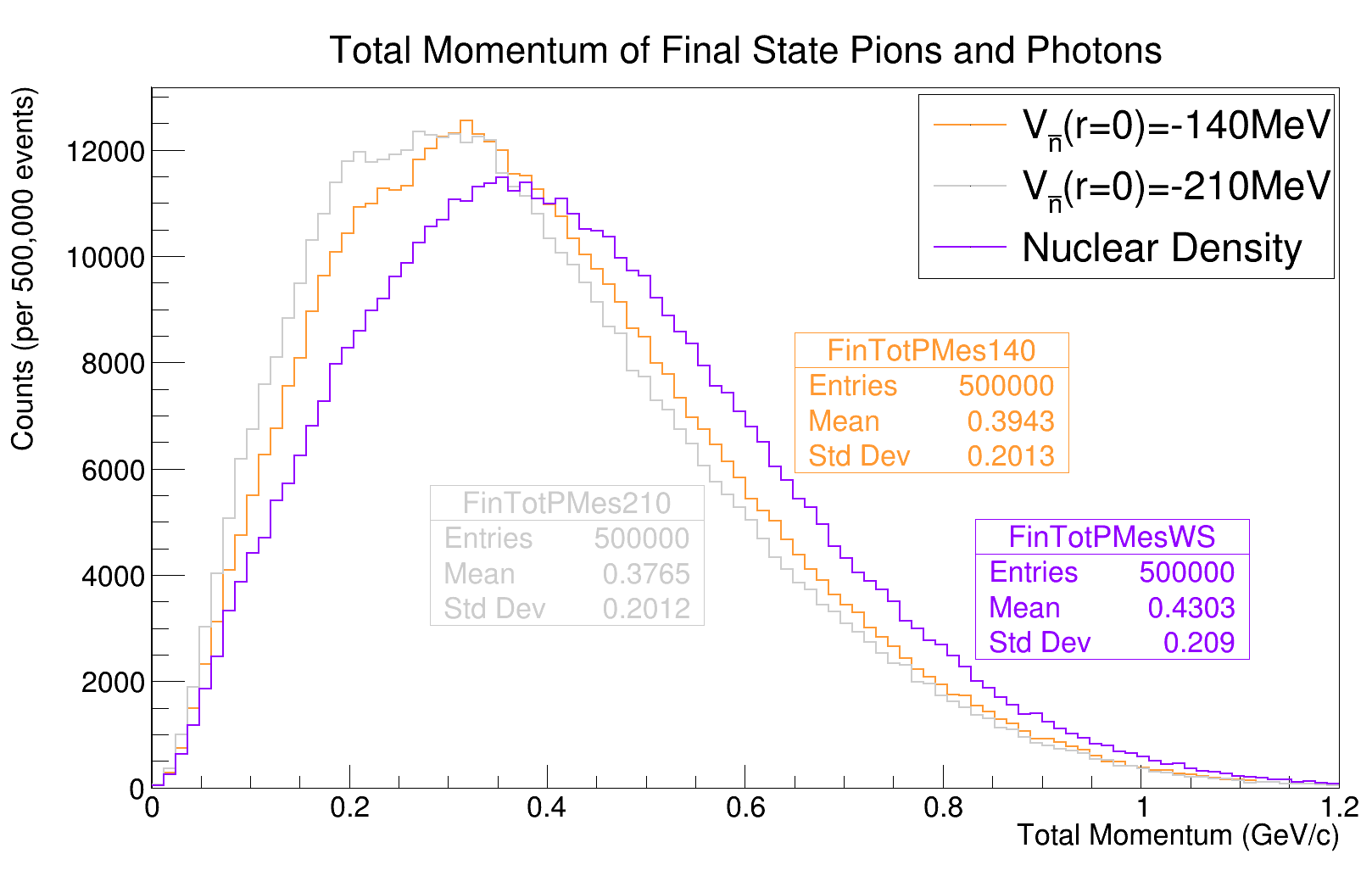}
    \includegraphics[width=0.49\columnwidth]{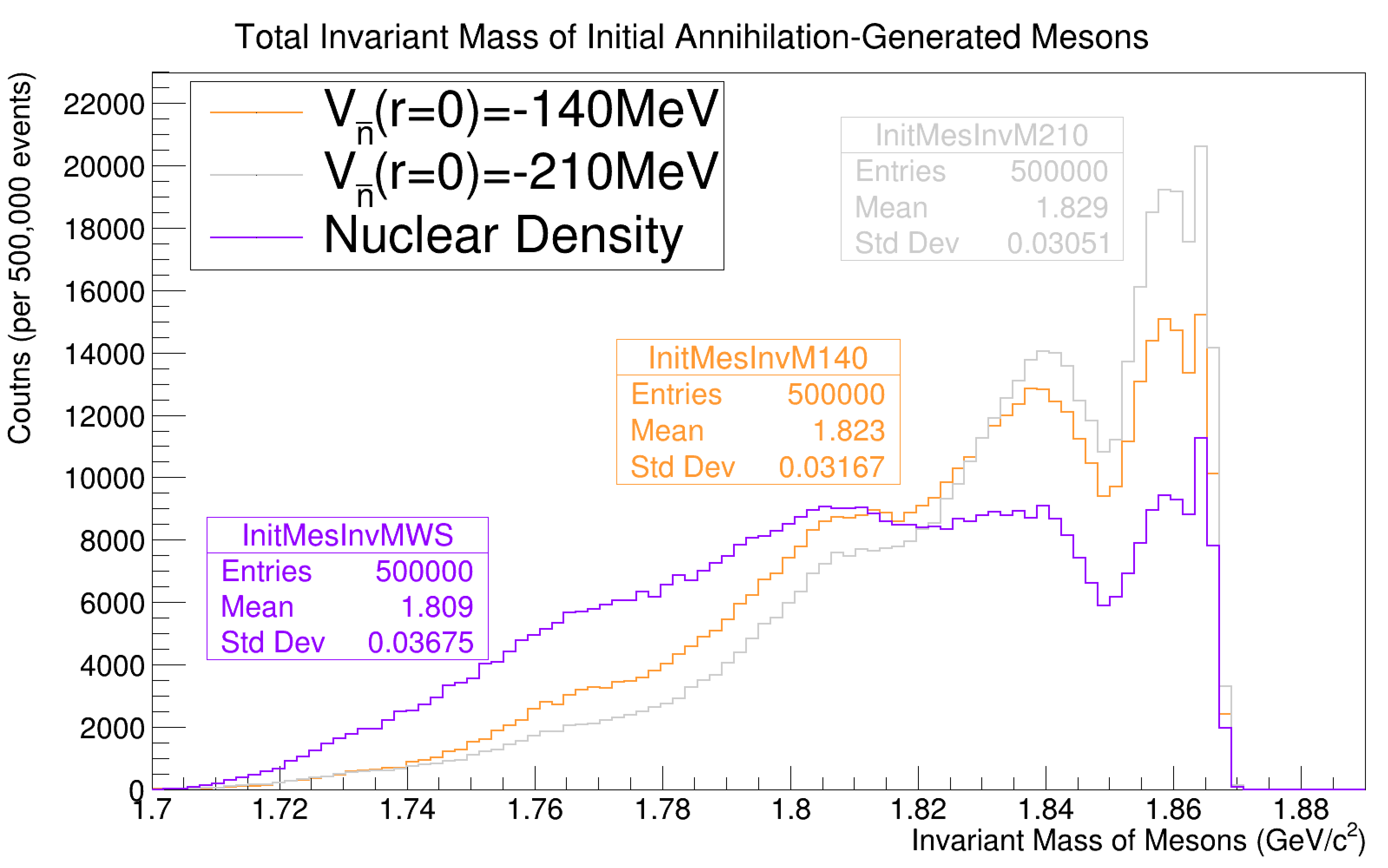}
    \includegraphics[width=0.49\columnwidth]{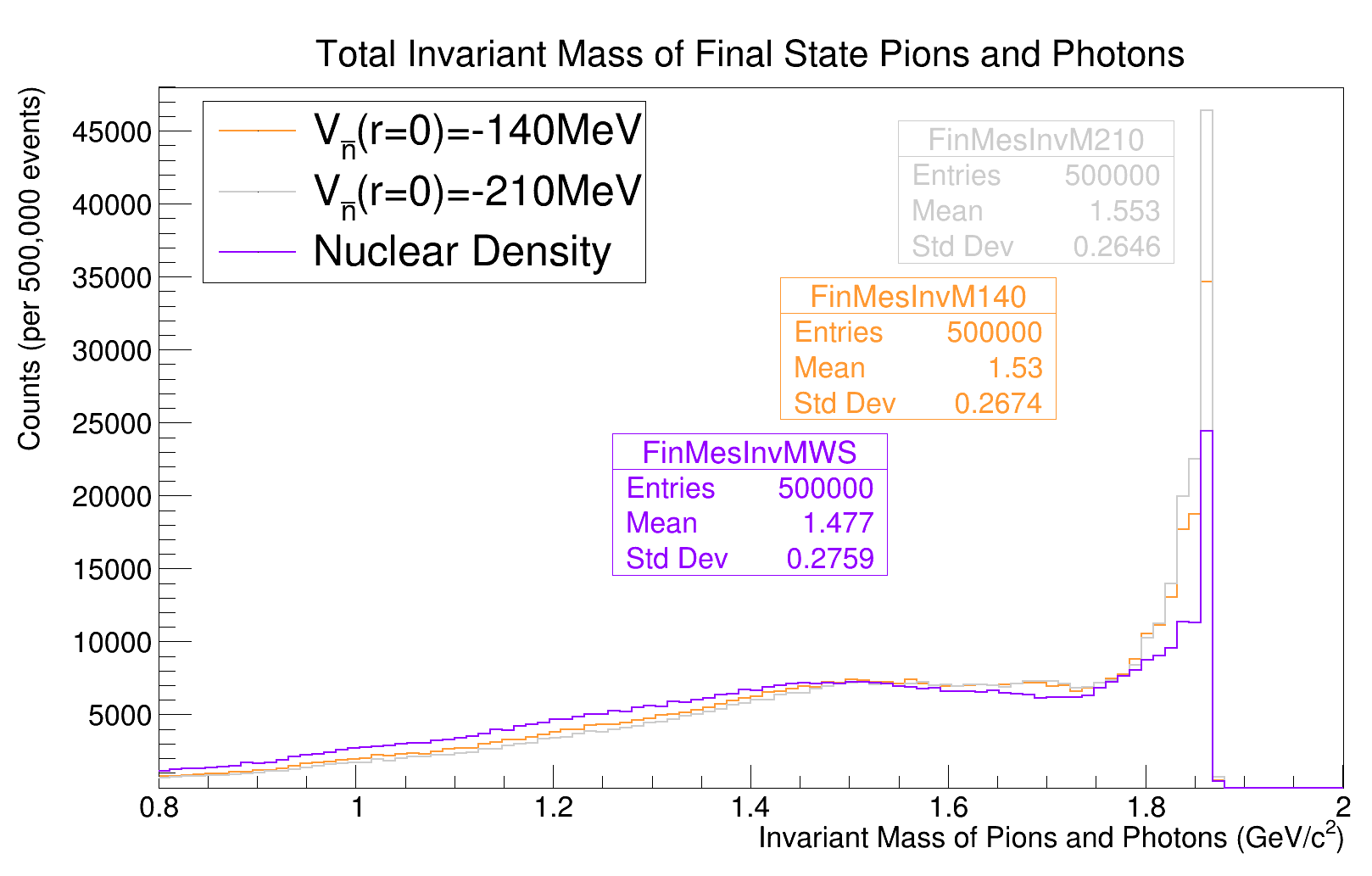}
    \caption{Top left: The initial total vector momentum of annihilation generated mesons is shown; by conservation, these distributions are equivalent for the initial annihilation pair. Top right: The final state total vector momentum of subsequent pions and photons is shown following heavy mesonic resonance decays and the intranuclear cascade. Bottom left: The initial total invariant mass of annihilation generated mesons is shown, where strengths of the expected peaks around $\sim 1.9\,$GeV critically depend on the strength of the antinucleon potential. Bottom right: The final state total invariant mass of subsequent pions and photons is shown following heavy mesonic resonance decays and the intranuclear cascade; due to rescattering and absorptive processes, many events move toward a lower valued tail. The hierarchy of peaks among the models remains, owing to the their progressively peripheral annihilations and thus avoidance of FSIs. All plots shown include the modern annihilation densities with $V_{\bar{n}}(r=0)=-140\,$MeV (orange), $V_{\bar{n}}=-210\,$MeV (grey), and an annihilation distribution derived from the nuclear density (purple); note that the model derived from the nuclear density also assumes $V_{\bar{n}}(r=0)=-140\,$MeV. Each is shown for 500,000 generated events.}
    \label{fig:FinAnnTotPInvMMes}
\end{figure}

\newpage

\twocolumngrid
Now consider the total vector momentum and invariant mass distributions of the mesonic annihilation products at the annihilation point and after leaving the nucleus following FSIs. The top left plot of Figs.~\ref{fig:FinAnnTotPInvMMes} shows the $\bar{n}N$ annihilation pair momentum distributions for all variants of the antinucleon potential and associated annihilation probability distributions available to the simulation, just as presented in Fig.~\ref{fig:AnnRadProbDistsO}; by conservation, this is exactly the initial total vector momentum distribution of all annihilation-generated mesons at the annihilation point, before FSIs. These distributions are simply the vector sum of $\bar{n}$ and $N$ species (see again Fig.~\ref{fig:InitNucMom}), and their high momentum tail depends on the potential $V_{\bar{n}}(r=0)$ depth. One might think that the greater the depth of the antinucleon potential, the greater the mass defect of the antinucleon, and thus the greater the momentum the annihilating pair can attain within the nucleus. However, as can be seen, this is not necessarily the case once the antinucleon potential grows to larger values, which actually leads to more peripheral annihilations and thus shifts to a lower total initial vector momentum. The bottom left plot of Figs.~\ref{fig:FinAnnTotPInvMMes} shows the distribution of the total invariant mass of annihilation products before FSIs for all simulation variants. At the initial annihilation stage, the result significantly depends on the annihilation radius, which, as discussed, is itself a function of the simulation parameters (Fig.~\ref{fig:AnnRadProbDistsO}). For example, the deeper the antinucleon potential, the further annihilation occurs at the periphery of the nucleus, and so one should expect more events with greater invariant mass. Thus, Fig.~\ref{fig:InitAnnE} and the left Figs.~\ref{fig:FinAnnTotPInvMMes} collectively demonstrate how the initial energy, momentum and invariant mass of the annihilation depend on the potential, a free parameter of the model. At the same time, when the dynamics of the annihilation process and associated modifications connected to the nuclear medium are taken into account in these simulations~\citep{Barrow:2019viz}, the shape of the spectra changes significantly. In this way, an important distinguishing feature of this work is the maintenance of correlation between the values of momentum and invariant mass with the annihilation radius.

Further, consider now how FSIs affect the expected distributions of observable momentum and invariant mass. It can be seen from the right plots of Figs.~\ref{fig:FinAnnTotPInvMMes} that FSI, even with the pronounced peripheral character of annihilation, significantly reduces the differences between the simulation options in all but the highest invariant mass region. This is due to the relative isotropy of the nuclear medium, wherein a large proportion of the particles emanating from the annihilation generated meson "stars" are still forced to move through the nucleus and undergo rescattering and absorption.

\begin{figure}[h!]
    \centering
    \includegraphics[width=1.0\columnwidth]{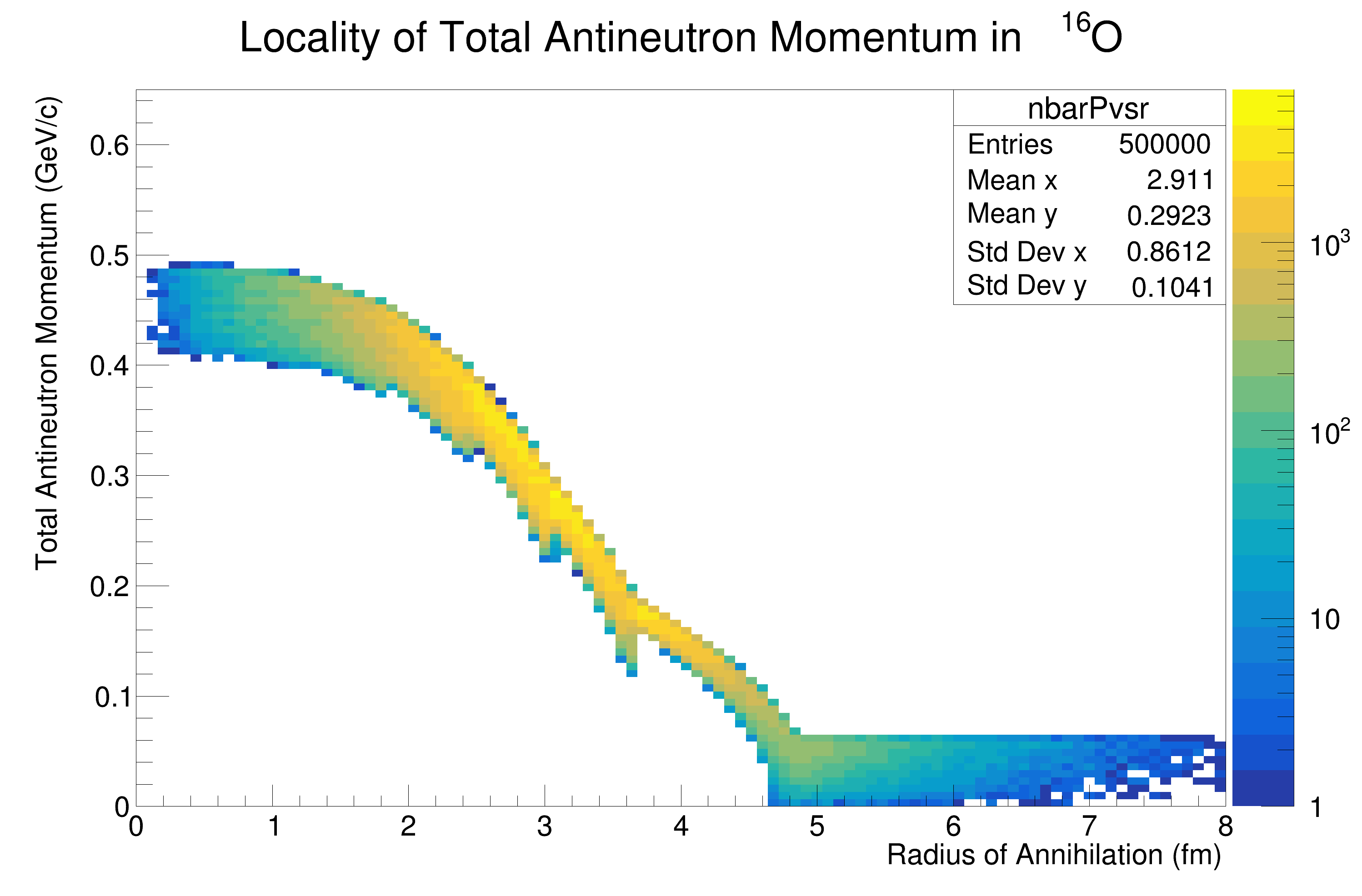}
    \includegraphics[width=1.0\columnwidth]{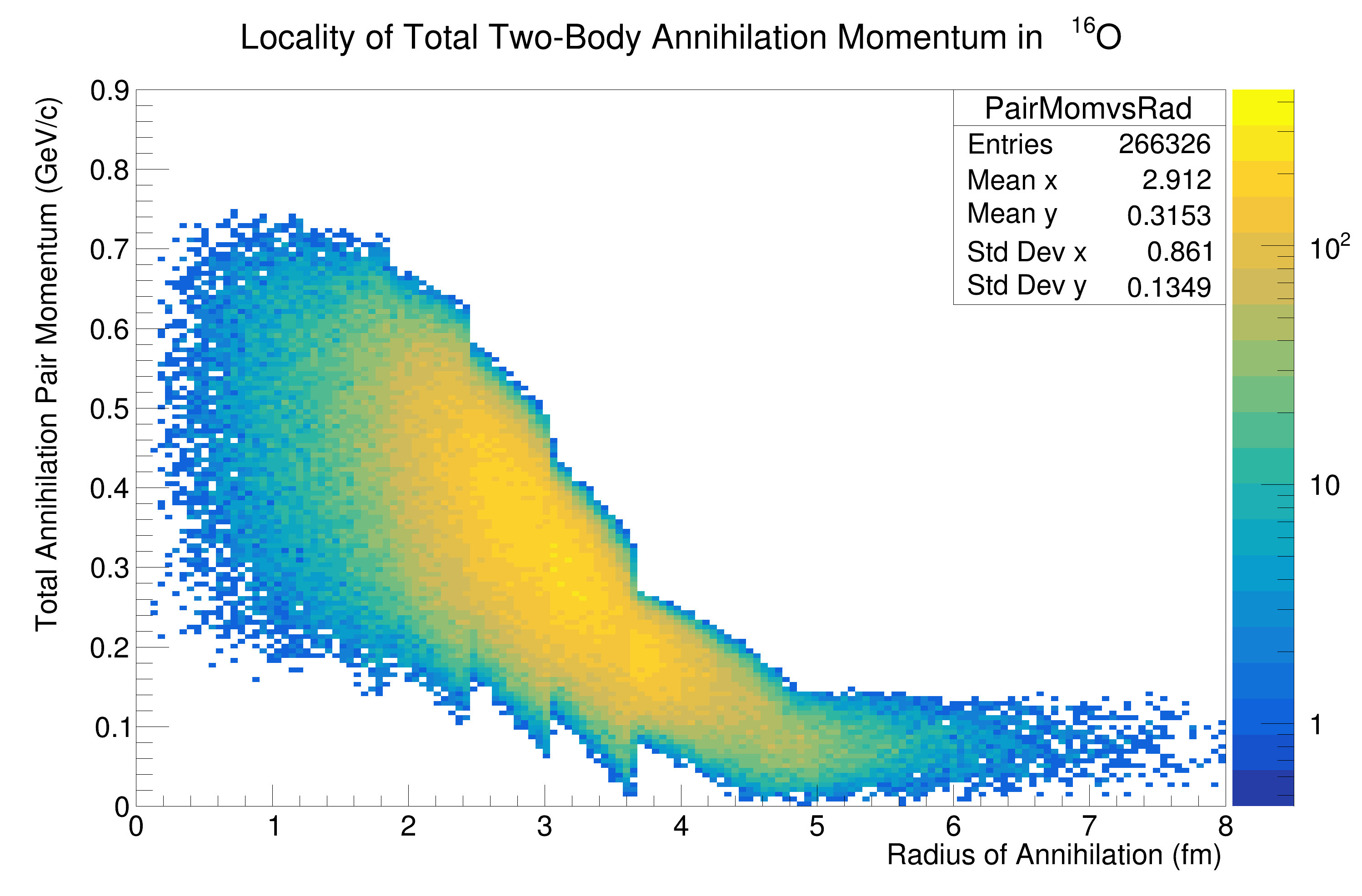}
    \caption{Top: The local nature of the $\bar{n}$ momentum is shown, with little to no visible zoned structure. Bottom: The local nature of the annihilation pair total vector momentum for intranuclear $\bar{n}p$ annihilation is shown. Each is shown for 500,000 generated events for an $\bar{n}$ potential of $V_{\bar{n}}(r=0)=-140\,$MeV.}
    \label{fig:NucMomvRad}
\end{figure}

A visual demonstration of the $\bar{n}$ and the $\bar{n}N$ annihilation pair's total momentum correlation with the annihilation radius is presented in Figs.~\ref{fig:NucMomvRad} for the variant of the calculation with $V_{\bar{n}}(r=0)=-140\,$MeV because from our point of view, the most reasonable value of the potential is $V_{\bar{n}}(r=0)=-140\,$MeV; thus, all following figures are presented for calculations utilizing this option. (see again Fig.~\ref{fig:AnnRadProbDistsO}).

\begin{figure}
    \centering
    \includegraphics[width=1.0\columnwidth]{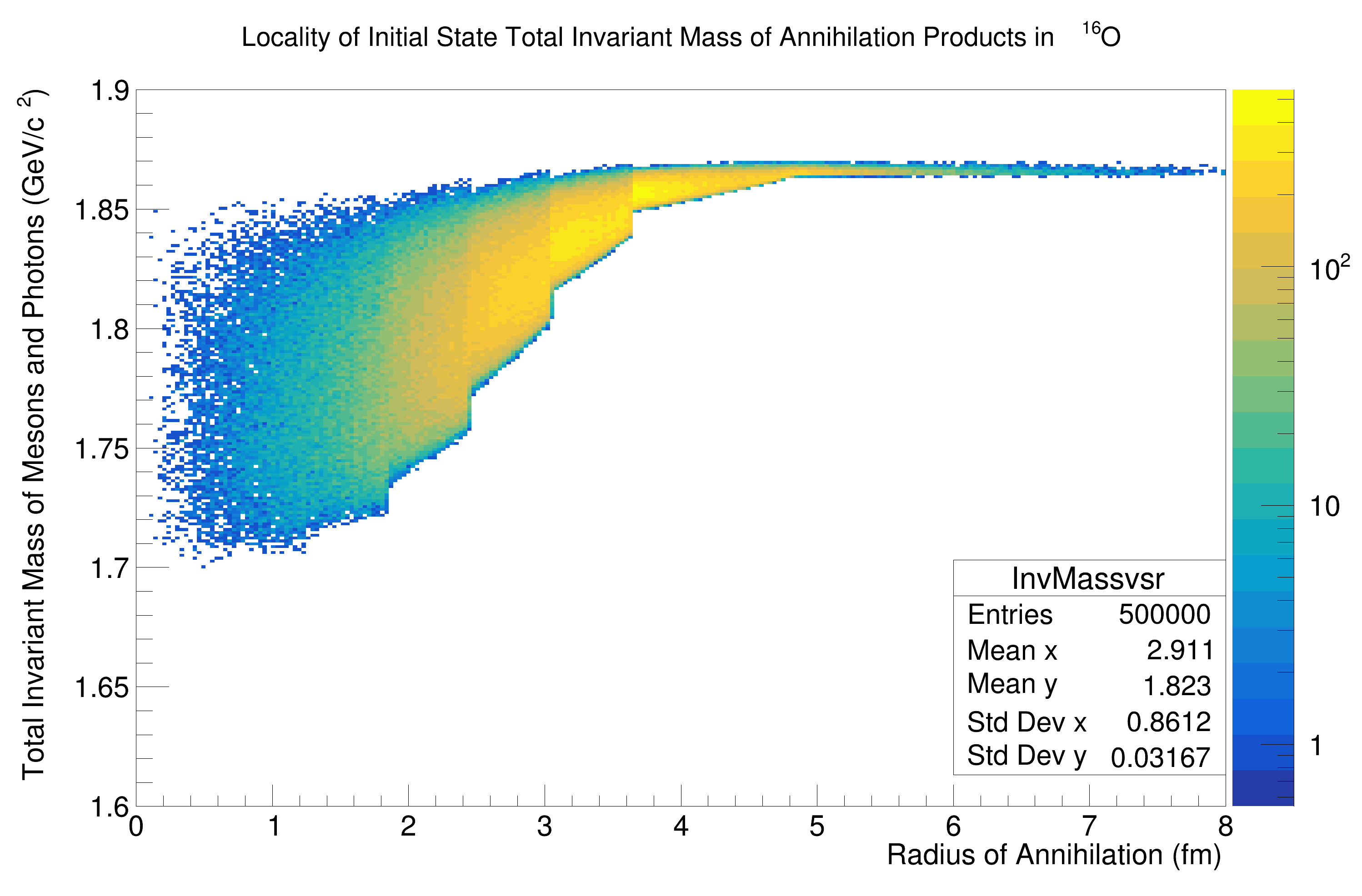}
    \includegraphics[width=1.0\columnwidth]{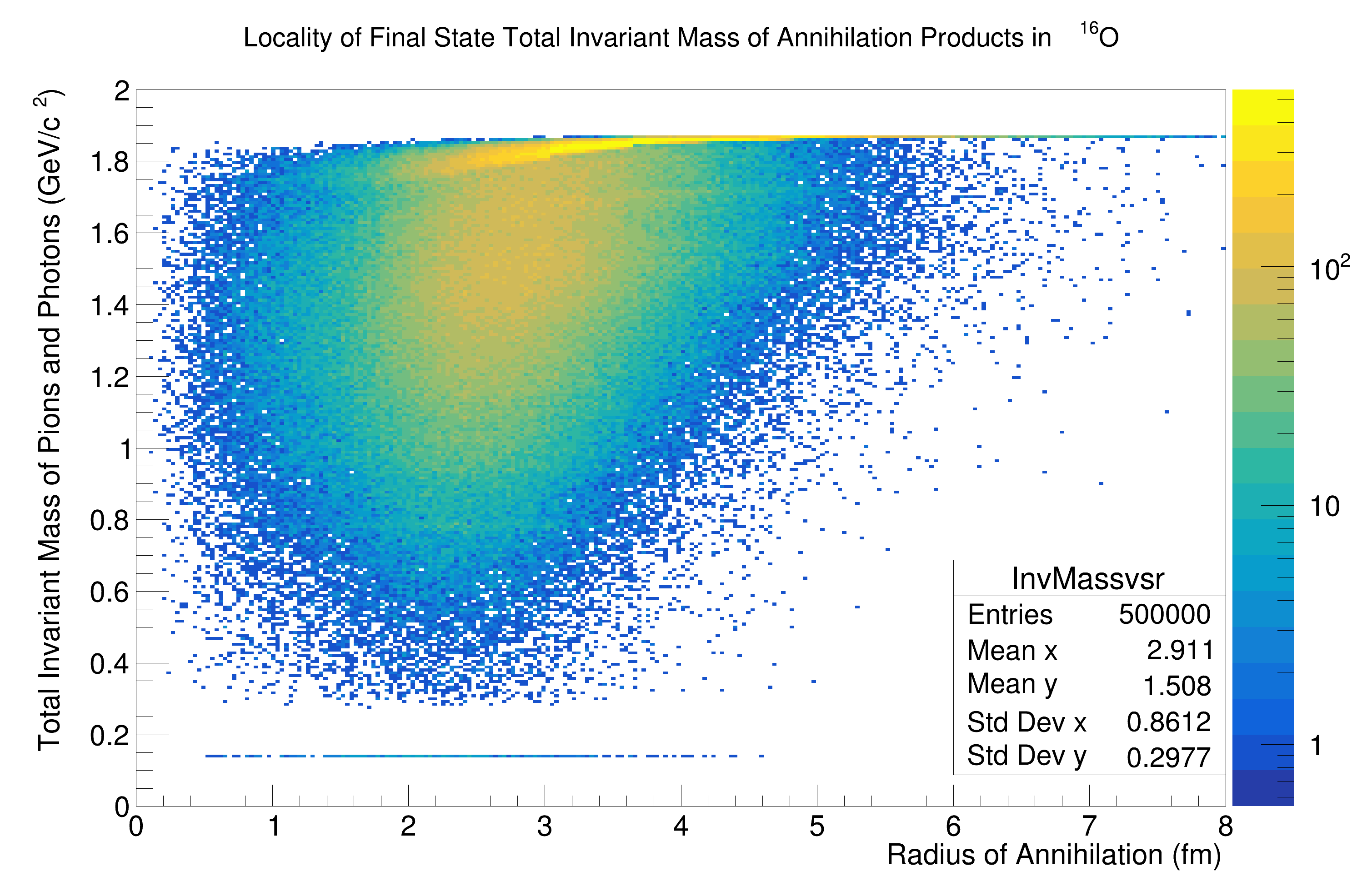}
    \caption{Top: The initial state's invariant mass of annihilation-generated mesons and photons vs. radius is seen before FSIs, showing the local effects of the (anti)nucleon potential and associated mass defects. Bottom: The same for the final state's truth invariant mass of all pions and photons following FSIs, showing the importance of taking account of both the (anti)nucleon potential and radial position of the annihilation to avoid excessive FSIs. Each is shown for 500,000 generated events for an $\bar{n}$ potential of $V_{\bar{n}}(r=0)=-140\,$MeV.}
    \label{fig:InvMassvsRad}
\end{figure}

Furthermore, in Figs.~\ref{fig:InvMassvsRad}, we see a parameter space showing the invariant mass of annihilation-generated mesons in the initial and final state as a function of annihilation radius. Critically, it is seen that a large proportion of the events are peripheral in character; more importantly, this is seen to drastically increase the predicted invariant mass in the final state (bottom). In contrast to the assumption of a uniform annihilation probability across the whole nucleus, due to the lessened FSIs, this simulation predicts a greater potential signal observation viability.

The final state, constructed following intranuclear transport of annihilation-generated pions and photons generated from heavy mesons decays as well as the de-excitation of all nuclear remnants, can also be investigated via iteration over the available model configurations. Figs.~\ref{fig:FinAnnTotPInvMMes} show the effect of increases in the antinucleon potential, where when combined with peripheral annihilation allow for a higher reconstructed invariant mass and similarly lower total vector momentum. Note that the invariant mass can decrease due to loss of particle, while this can thus increase (cause greater imbalance) to the total vector momentum. Though some of these differences appear marginal in one dimension, the local correlations of these variables together can greatly affect the observability of the signal.

\begin{figure}[h!]
    \centering
    \includegraphics[width=1.0\columnwidth]{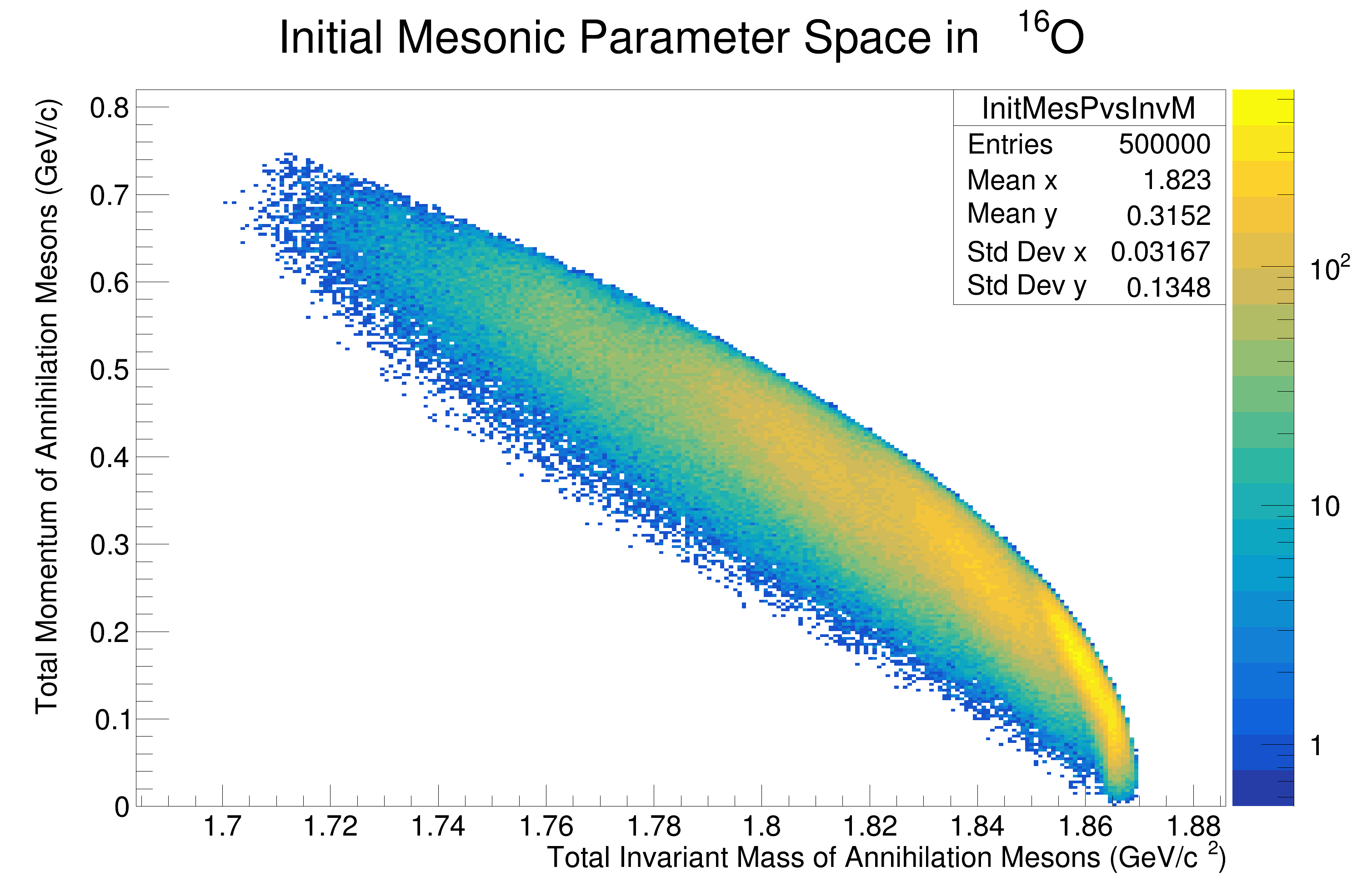}
    \includegraphics[width=1.0\columnwidth]{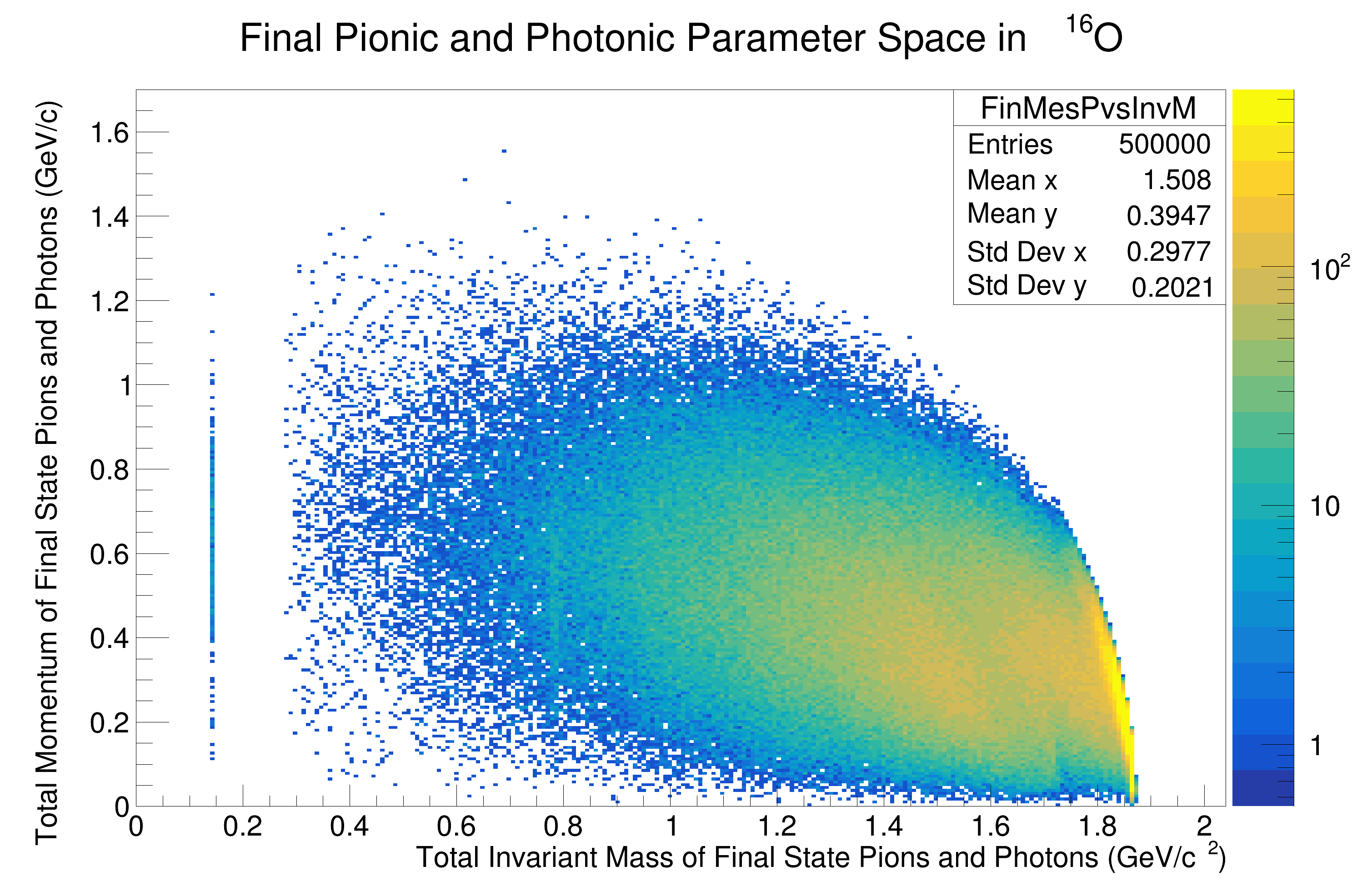}
    \caption{The initial and final state mesonic parameter space is shown for all mesons and photons generated from the annihilation. Top: Decrease of invariant mass is due to the off-shell nature of the annihilating (anti)nucleons. Bottom: Final state interactions cause the rescattering of and losses of mesons, generating a less constrained parameter space. Each is shown for 500,000 generated events for an $\bar{n}$ potential of $V_{\bar{n}}(r=0)=-140\,$MeV.}
    \label{fig:mesparamspace}
\end{figure}

The mesonic parameter space (total momentum vs. total invariant mass of annihilation-generated mesons) is informative of the initial annihilation dynamics and the effects of final state interactions on the signal, and is presented in Figs.~\ref{fig:mesparamspace}. In the top plot of Figs.~\ref{fig:mesparamspace}, the initial condition of the annihilation generated mesons is shown before intranuclear transport; here, the invariant mass decreases due to of-shell mass defects in correlation with radial position. The bottom plot of Figs.~\ref{fig:mesparamspace} shows the same parameter space, though now after the mesons' intranuclear propagation and fast decays of $\rho$, $\omega$, and $\eta$ mesons. Note that this model includes photons in the final state from $\eta$ and $\omega$ resonance decays. The disconnected regions toward the left of the plots are signs of single pion emission after at least one or more meson absorptions.

\begin{figure}[h!]
    \centering
    \includegraphics[width=1.0\columnwidth]{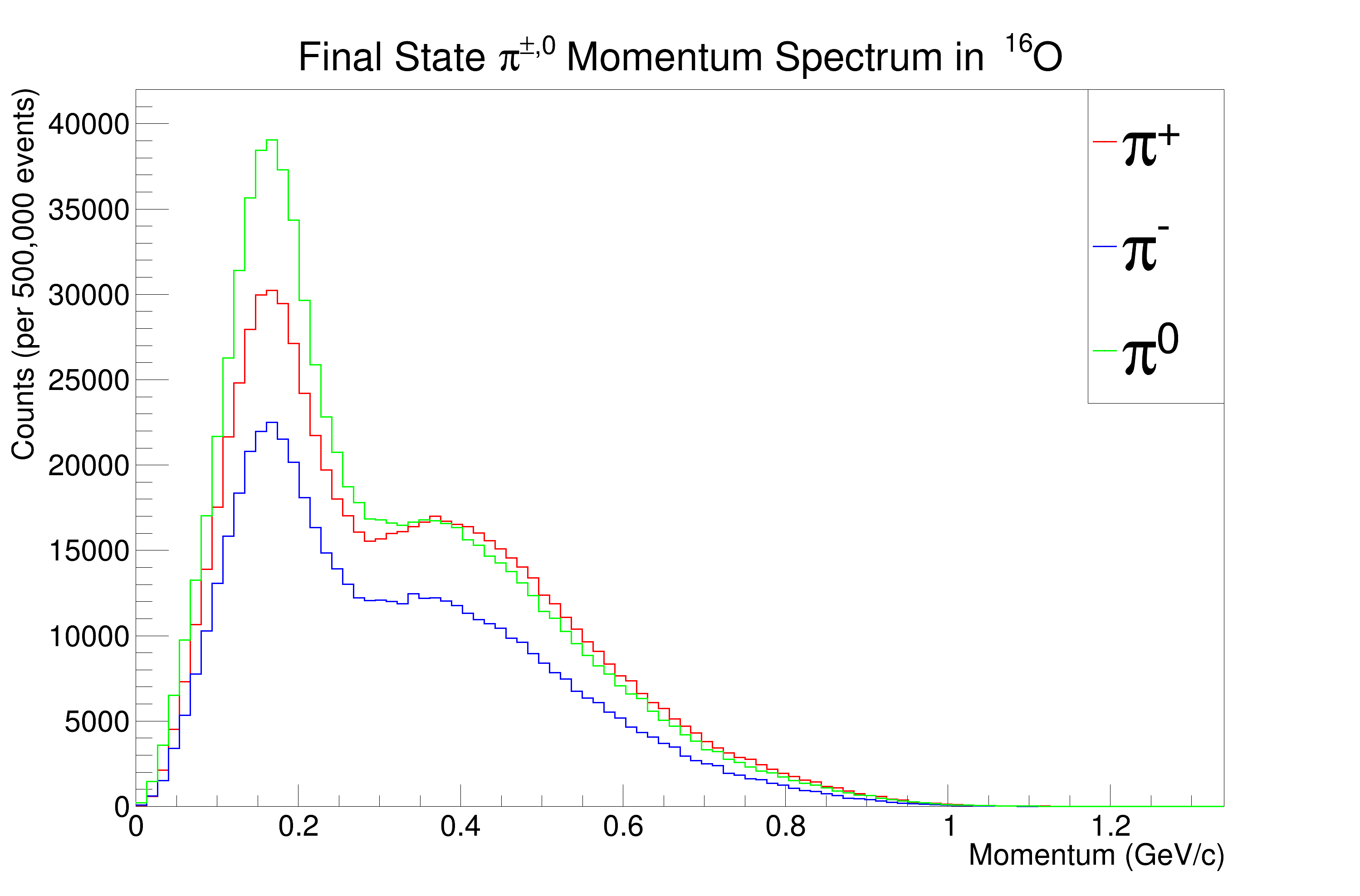}
    \includegraphics[width=1.0\columnwidth]{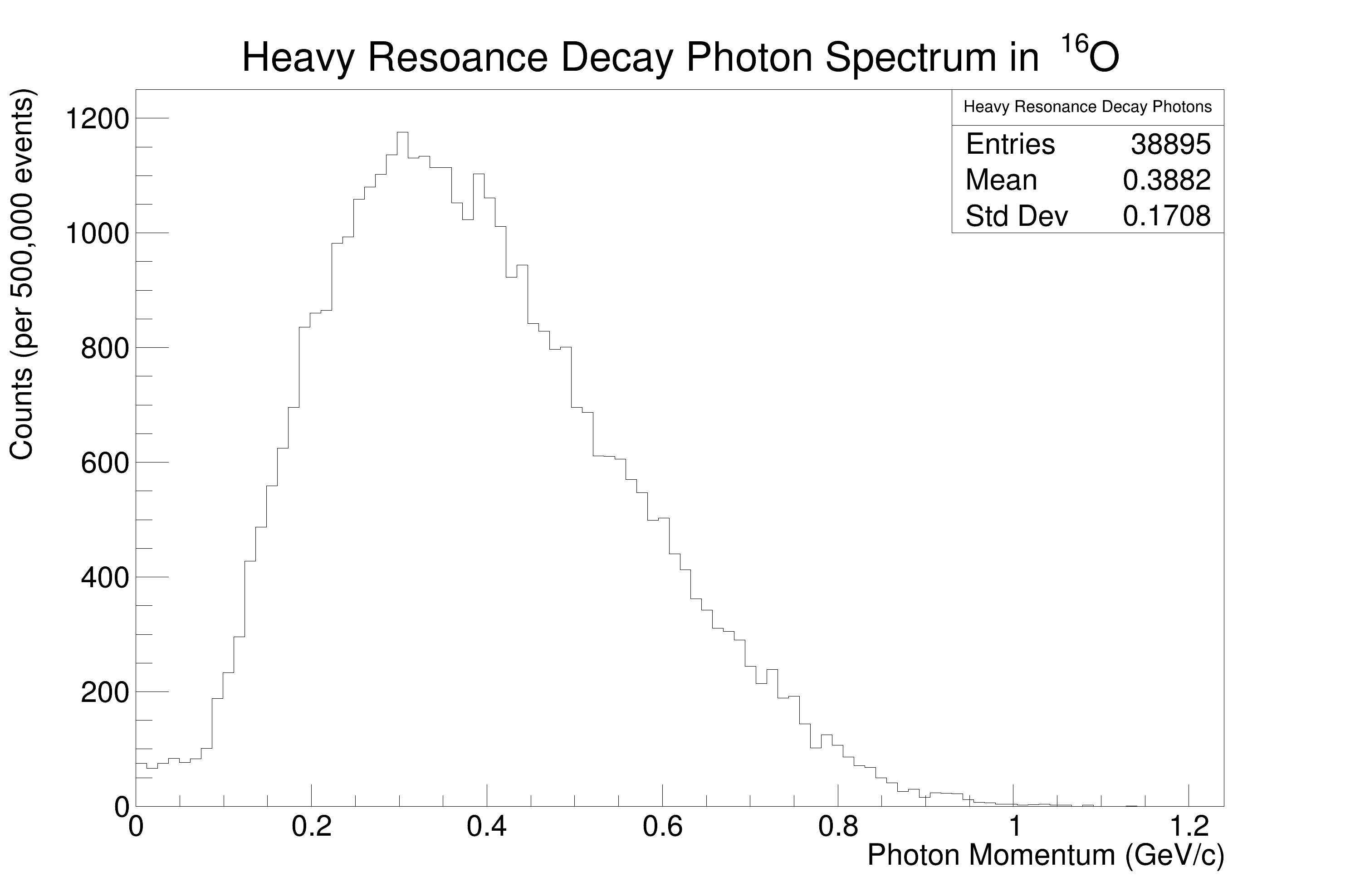}
    \includegraphics[width=1.0\columnwidth]{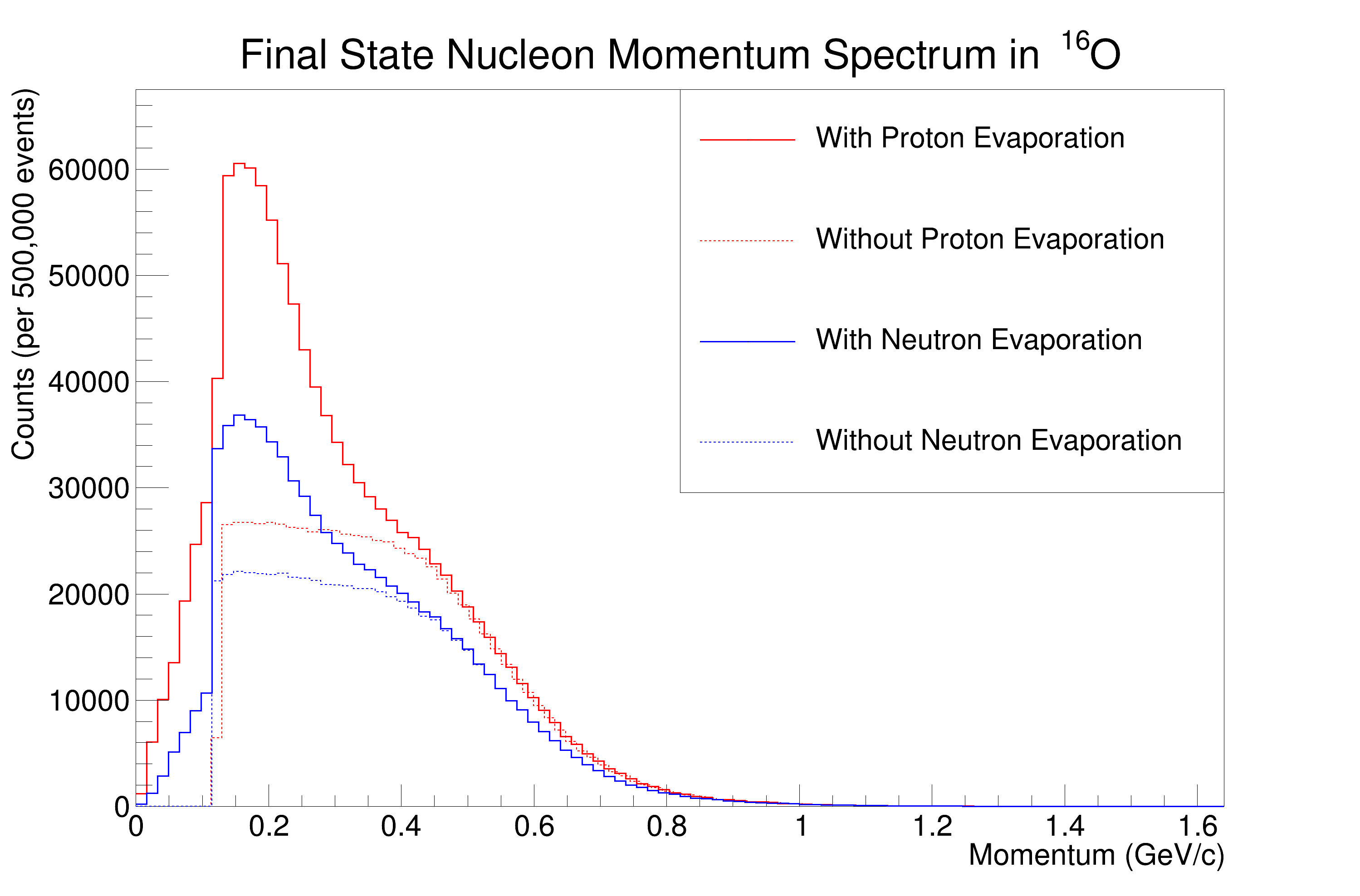}
    \caption{Top: Momentum spectra for all final state pion species. Middle: Heavy mesonic resonances can arise following an $\bar{n}N$ annihilation within the nucleus, some of which  may decay into photons; see~\citep{Golubeva:2018mrz} for branching fractions. These occur for only $\lesssim 10\%$ of events. Bottom: Final state momentum spectra for neutrons (blue) and protons (red), shown with (solid) and without (dashed) the consideration of fragmentary and evaporative processes.}
    \label{fig:MesDecPhotonsNucs}
\end{figure}

In Figs.~\ref{fig:MesDecPhotonsNucs} the pions' (top), resonance decay photons' (middle), and final state nucleons' (bottom) momentum spectra are shown. Note here that the decay photons arise primarily via processes such as $\eta \rightarrow 2\gamma$ with a branching fraction of $39.3\%$, $\eta \rightarrow \pi^+\pi^-\gamma$ at $4.9\%$, and $\omega\rightarrow\pi^0\gamma$ at $8.7\%$~\citep{Golubeva:2018mrz}.

\begin{table*}
    \caption{A list of final state particle multiplicities from three versions of the model for intranuclear $\bar{n}$\isotope[15][8]{O}, taking into account all annihilation branching ratios, the intranuclear antinucleon potential, and an associated nuclear medium response~\citep{Golubeva:2018mrz,Barrow:2019viz}. Based on simulations of 500,000 events with different intranuclear antinucleon potential depths. Nucleon multiplicities include evaporative processes.}
    \label{tab:oxygen_multiplicities}
    \centering
    \begin{ruledtabular}
    \begin{tabular}{c|cccccccc}
         Simulation & $M(\pi)$ & $M(\pi^+)$ & $M(\pi^-)$ & $M(\pi^0)$ & $M(\gamma^\text{res.}_\text{dec.})$ & $M(\gamma^\text{nuc.}_\text{rem.})$ & $M(p)$ & $M(n)$ \\
         \hline
         $\bar{n}$\isotope[15][8]{O} w/~$V_{\bar{n}}=|210|\,$MeV & $4.37$ & $1.56$ & $1.12$ & $1.68$ & $0.08$ & $0.34$ & $2.14$ & $1.44$ \\
         $\bar{n}$\isotope[15][8]{O} w/~$V_{\bar{n}}=|140|\,$MeV & $4.33$ & $1.54$ & $1.10$ & $1.68$ & $0.08$ & $0.31$ & $2.28$ & $1.54$ \\
         $\bar{n}$\isotope[15][8]{O} w/Nuclear Density           & $4.23$ & $1.49$ & $1.07$ & $1.67$ & $0.07$ & $0.27$ & $2.59$ & $1.78$ \\
    \end{tabular}
\end{ruledtabular}
\end{table*}

All of these plots are well summarized in Table~\ref{tab:oxygen_multiplicities}, which shows the average multiplicities of pions, photons, and nucleons emitted by the oxygen nucleus as a result of the $n\rightarrow\bar{n}$ intranuclear transition. It can be seen that the closer the annihilation occurs to the center of the nucleus, the more significant the role FSIs play via the absorption of pions and, accordingly, the smaller the number of pions which exit the nucleus; correspondingly, this creates a higher number of nucleon knock-outs. The similar multiplicity of heavy-resonance decay photons across the model configurations is due to limited rescattering of these short-lived species and lack of photon absorption processes within the intranuclear cascade. A progressively decreasing de-excitation photon multiplicity is observed as the annihilation position becomes more interior; this is due to the increasingly violent breakup of the nucleus, reducing the overall number of potential nuclear remnants with $A\geq2$ and excitation energies conducive to photon emission.

Each of these can have critical implications for any definitive observation of an intranuclear $n\rightarrow\bar{n}$ event in~\isotope[16][8]{O}. Also, the three models presented (with different values of the intranuclear $\bar{n}$ potential $V_{\bar{n}}$~\citep{Barrow:2019viz}) can help to estimate potential uncertainties in the event generator beyond those discussed within Sec.~\ref{sec:review} and~\citep{Golubeva:2018mrz,Barrow:2019viz}.

\section{Implications}
Through the above discussions, one comes to realize the important interplay between a few key variables: the strength of the antinucleon potential, the annihilation radius, along with the total momentum and the total invariant mass of initial and final state mesons and photons. Though the annihilation potential and radius cannot be determined from what may in the end be but a single observed event, the dependence of the two other definitive observables on these variables is critical. Namely, even when accounting for quantum effects which produce a more peripheral annihilation, a stronger antinucleon potential together yields still fewer final state interactions, thus increasing the possibility of reconstructing a pionic-photonic system with higher invariant mass and lower total vector momentum; this makes the observation of an $n\rightarrow\bar{n}$ event more probable in that it is likely to occur within a more localized (and arguably more reconstruction-stable) part of the mesonic parameter space. Ignoring these important correlations can thus act to limit the final experimental lower limit sensitivities one extrapolates from current measurements, even when overcome by irreducible atmospheric neutrino backgrounds. The qualia of these observables become still more important as the field adopts more automated techniques, such as with machine learning~\citep{Hewes:2017xtr,Barrow:2021odz,YeonJaeThesis}.


\section{Future Work}
This set of improvements to the underlying antineutron-annihilation model can be applied to other nuclei, including for nuclei relevant for the Deep Underground Neutrino Experiment ($^{40}$Ar) and the European Spallation Source's HIBEAM/NNBAR program ($^{12}$C). The addition of de-excitation photons to this model add extra potentially observable qualities to the expected signal, possibly cutting background still further. Secondarily, given the modularity of the new final state interactions code, it may be possible to apply such computational techniques other types of interactions via their generated initial state four-momenta, allowing for more consistent comparisons between various signals and backgrounds, such as atmospheric neutrinos and cosmogenic muons. There is interest in extending this work toward modeling of proton decay inside a nucleus. 

\section{Conclusions}
A new set of simulations of intranuclear $\bar{n}$ annihilation within the \isotope[16][8]{O} nucleus have been completed with the addition of a modern statistical nuclear disintegration model capable of predicting the de-excitation photon spectra of nuclear remnants. While the addition of these de-excitation photons and nuclear remnants to the model make it more physically complete and in principle create still more handles to discriminate signal from background, this will likely prove difficult within water Cherenkov detectors given their high hadronic thresholds and expected $\pi^0\rightarrow\gamma\gamma$ decays dominating the detector. The nuclear disintegration technique is portable to other nuclei of interest, and it is hoped that such additional realistic outputs from the model will permit more sensitive searches for $n\rightarrow\bar{n}$ transformations at current and future experiments such as Super- and Hyper-Kamiokande, NNBAR at the European Spallation Source, and the Deep Underground Neutrino Experiment. Further, a modern quantum mechanical radial $\bar{n}$ annihilation probability distribution has been calculated for the \isotope[16][8]{O} nucleus, predicting an associated intranuclear suppression factor of $T_{R}=0.65\times 10^{23}\,\mathrm{s}^{-1}$, in line with past estimates. This annihilation probability distribution has been integrated into the intranuclear annihilation simulation framework, allowing for prediction of novel final states of annihilation-generated pions. Critical among the findings of this work is the important interplay of the initial annihilation position within the nucleus depending on the strength of the antinucleon potential, and from this the final states pions' observable total momentum and invariant mass. It is likely that the peripheral character of the annihilation will increase potential sensitivities to dinucleon decay signals, as fewer final state interactions occur when these correlations are properly accounted for. Samples of 500,000 events or more are available upon request to the authors.

\section{Acknowledgements}
JLB's initial work on this project was partially supported by the Visiting Scholars Award Program of the Universities Research Association. Any opinions, findings, and conclusions or recommendations expressed in this material are those of the authors and do not necessarily reflect the views of the Universities Research Association, Inc. JLB would also like to thank The University of Tennessee at Knoxville Department of Physics and Astronomy, as well as the Zuckerman Institute, for partial support of this work at various later stages. JLB is a Zuckerman Postdoctoral Scholar. ASB acknowledges the support of RFBR (Russia) of this work through research Project No. 18-02-40084.

\clearpage
\bibliography{bibliography}

\end{document}